\documentclass[prb,twocolumn,showpacs,amsmath,amssymb]{revtex4-1}

\usepackage[dvipdfmx]{graphicx}

\usepackage{bm}

\usepackage{color}
\usepackage{braket}
\usepackage{physics}
\usepackage{ulem}

\begin{document}

\title{Unusual wave-packet spreading and entanglement dynamics in non-Hermitian disordered many-body systems}

\author{Takahiro Orito}
\author{Ken-Ichiro Imura}
\affiliation{Graduate School of Advanced Science and Engineering, Hiroshima University, 739-8530, Japan}

\date{\today}

\begin{abstract}

Non-Hermiticity and dephasing, collaborating
in a unusual wave packet dynamics,
realizes unconventional entanglement evolution
in a disordered, interacting 
and 
asymmetric (non-reciprocal) quantum media.
Taking the Hatano-Nelson model as a concrete example, 
we first consider how wave packet spreads in a non-Hermitian disordered system 
for demonstraing 
that it is very different from the Hermitian case. 
Interestingly, a cascade like wave packet spreading 
as in the Hermitian case is suppressed in the clean limit and at weak disorder, 
while it revives in the vicinity of the localization-delocalization transition.
Based on this observation, we then analyze how the entanglement entropy 
of the system evolves in the interacting non-Hermitian model, 
revealing its non-monotonic evolution in time.
We clarify the different roles of dephasing in the time evolution of entanglement entropy
in Hermitian and non-Hermitian systems, and show that
the many-body dynamics 
is governed by a principle 
different from the Hermitian case. 
The size dependence of the results suggests with the increase of disorder,
a unusual area-volume-area law 
crossover of the maximal entanglement entropy.
To analyze the effects of disorder on a firm basis, using the Hermitian limit as a benchmark,
we employ a quasi-periodic disorder (Aubry-Andr\'{e} model)
in the analyses.
\end{abstract}

\maketitle

\section{Introduction}


The entanglement entropy 
has proven to be one of
the useful measures
for characterizing many-body localization (MBL)
\cite{MBL1,MBL2}
both in its static 
\cite{Nayak,MBLent5,area_law2,MBLent1,MBLent2}
and dynamical properties.\cite{log3,BPM,Abanin_log,Trans1,Trans2,Trans3}
The MBL phase appears quite generically;
i.e., both in theories\cite{MBL3,MBL4,MBL5} and in experiments\cite{MBL6,MBL7,MBL8,MBL9,MBL10},
in disordered (interacting) many-body systems,
constituting a counterexample to the
eigenstate thermalization hypothesis (ETH),\cite{ETH1,ETH2,ETH3}
the quantum mechanical version of ergodicity.
Disorder tends to localize the wave function; cf. Anderson localization
in the non-interacting limit,
while
here in an interacting system the so-called
quasi local integrals of motion (LIOMs)
\cite{LIOM1,LIOM2}
play the roles of localized wave functions in an Anderson insulator.
Anderson localization\cite{AndersonL} occurs in real space, while
MBL manifests in Fock space\cite{FockLB,FockLL1,FockLL2,FockLO}
due to the
emergence of LIOMs.
Many-body wave functions in the MBL phase take the form of a simple product
in the basis of LIOMs,\cite{LIOM1,FockLO}
leading to a strong suppression of entanglement entropy;
i.e.,
the area-law behavior,
\cite{Nayak,MBLent5,area_law2}
manifesting a sharp contrast to the standard (ergodic)
volume law
in the ETH phase.

In addition to such a volume-to-area law crossover at the ETH-MBL transition,
the entanglement entropy also shows a unique dynamics in the MBL phase;
\cite{log3,BPM,Abanin_log}
in the quench dynamics
the entanglement entropy stays 
constant on average after an initial growth
in the non-interacting case, 
[see e.g., Fig. \ref{fig_EE_V0}, panel (i-a)]
while in the MBL phase,
it shows a peculiarly slow
(typically logarithmic in time)
but unbounded
growth in the MBL phase (case of $V\neq 0$),
reflecting 
the fact the wave functions (LIOMs) are exponentially localized
but the information can still flow;
\cite{Abanin_log,Huse_FMBL}
[see e.g., Fig. \ref{fig_EE_V}, panel (i-a)].
The initial growth 
is mainly due to the increase of the number entropy $S_{\rm num}$,
while
the subsequent logarithmic growth
stems 
the configuration entropy $S_{\rm conf}$
[see Eqs. (\ref{S_num}) and (\ref{S_conf}) for definitions of $S_{\rm num}$ and $S_{\rm conf}$;
cf. Sec. \ref{S_numconf} for details].
Physically,
the increase of $S_{\rm num}$ is due to spreading of the wave packet,
while 
dephasing also contributes to the increase of $S_{\rm conf}$.
In spreading of the wave packet 
the focus is on the amplitude of the wave function,
while
in dephasing
it is on the phase of the wave function.
To be precise,
dephasing occurs only in interacting many-body systems.
Also, $S_{\rm num}$ is local,
while $S_{\rm conf}$ is non-local in real space.
Thus, the entanglement entropy $S_{\rm tot}=S_{\rm num}+S_{\rm conf}$ measures both
(i) the relaxation of the initial state due to wave packet spreading, and
(ii) the growth of quantum correlation in the process of dephasing.

So far we have in mind 
the standard Hermitian ETH-MBL systems,
while
the behavior of entanglement entropy 
in non-Hermitian systems with non-reciprocal (or asymmetric) 
hopping 
\cite{Hatano1,Hatano2}
$\Gamma_L=e^g \Gamma_0$ vs.
$\Gamma_R=e^{-g} \Gamma_0$
[$g$ is the degree of non-reciprocity; see also
Eqs. (\ref{ham_sp}), (\ref{ham_mp})],
has been
also of some interest recently.
It has been reported 
in Refs. \onlinecite{Hamaz,china,panda}
that
the entanglement entropy $S_{\rm tot}(t)$ shows a non-monotonic time evolution;
i.e., at some specific parameter setting
the entanglement entropy $S_{\rm tot}(t)$ turns to decrease after its initial growth.
Such a non-monotonic behavior of $S_{\rm tot}(t)$
has been attributed
to the presence of a finite imaginary part 
in the eigenvalues $E$
which typically appears in this system;
when ${\rm\ Im} E \neq 0$,
the wave function is extended
and is sensitive to the non-reciprocality of the hopping.
On contrary,
in the MBL phase
the entanglement entropy $S_{\rm tot}(t)$ 
converges to a single value after some duration.
\cite{Hamaz}
In spite of these preceding studies
(Refs. \onlinecite{Hamaz,china,panda})
there still remains some open issues;
a comprehensive understanding of the various behaviors of $S_{\rm tot}(t)$
at different parameter settings
seem to be lacking.
Here, in this work we perform a systematic study of
$S_{\rm tot}(t)$
in non-Hermitian systems with non-reciprocal hopping,
and reveals the nature of its intricate behaviors 
that differ significantly from those in Hermitian systems.
We point out that
another factor
underlying
the difference in the behaviors of $S_{\rm tot}(t)$
in the Hermitian and non-Hermitian systems
is a very different way 
how a wave packet spreads in the two systems.
\begin{figure}
\includegraphics[width=85mm]{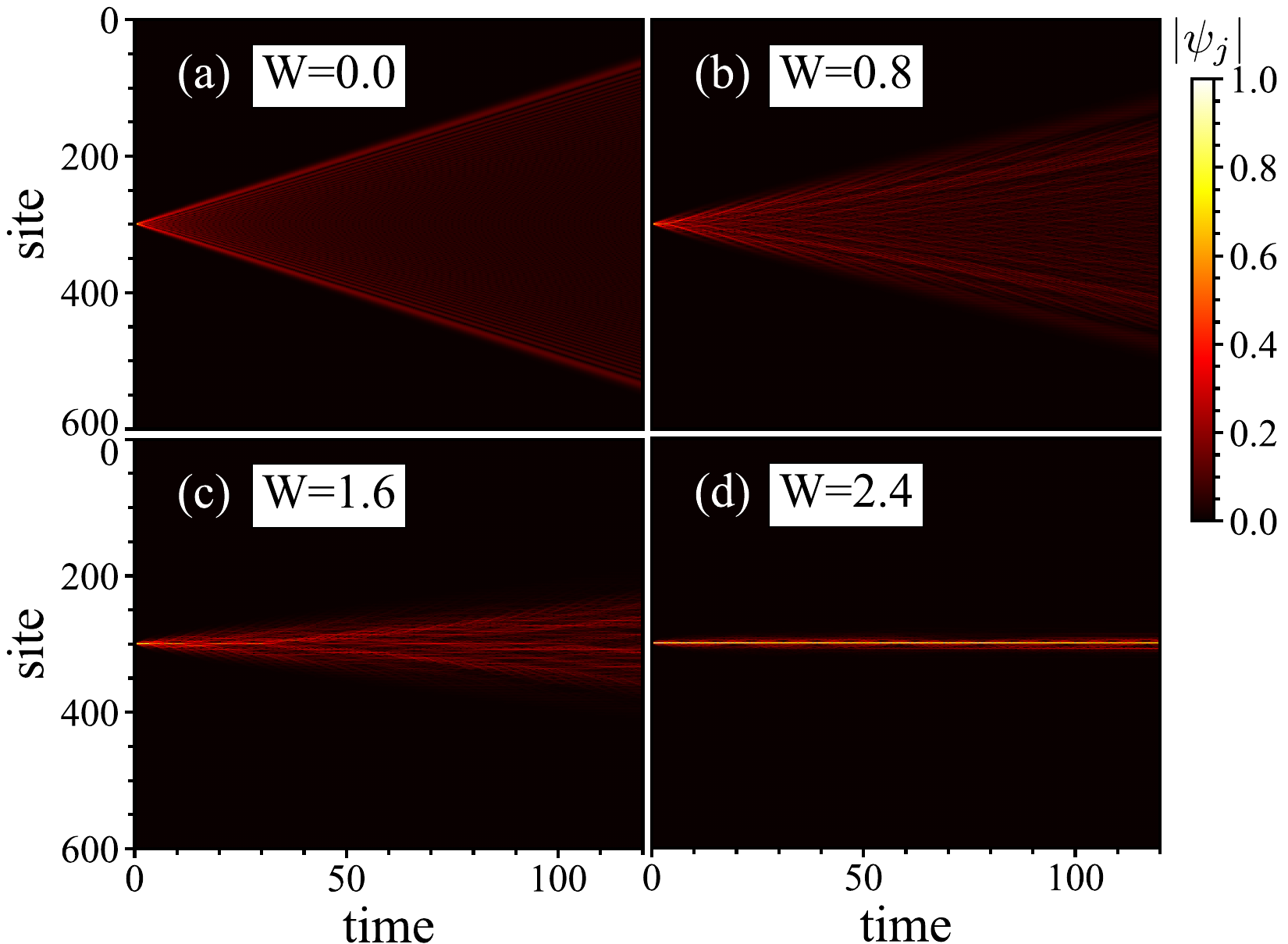}
\caption{Single-particle dynamics in the Hermitian case: $g=0$. 
Time evolution of the initial wave packet:
$\psi_j (0)=\delta_{j,j_0}$
is visualized.
The amplitude $|\psi_j (t)|$ [cf. Eq. (\ref{psi_t})]
is shown
by a gradation of plot colors indicated in the color bar.
The abscissa represents time $t$, 
and the ordinate the site $j$.
Different panels (a-d) correspond to different values of disorder strength $W$;
$W=0.0, 0.8, 1.6, 2.4$, respectively, for panels (a-d).
$\theta_0$ is fixed at $\theta_0=0$. No disorder averaging.
$j_0=300$, $L=601$.
}
\label{fig_1herm}
\end{figure}
\begin{figure}
\includegraphics[width=85mm]{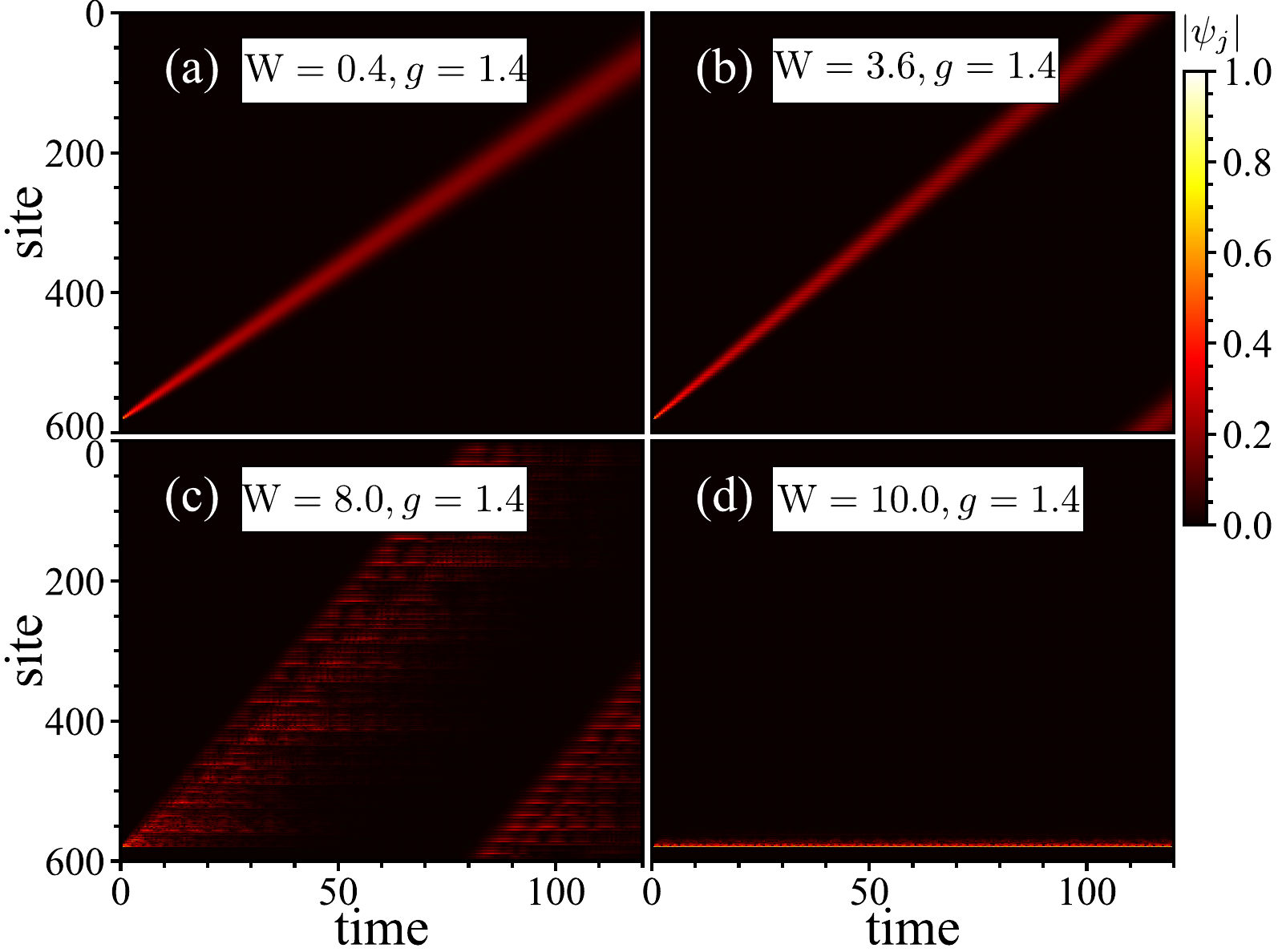}
\caption{Single-particle dynamics in the non-Hermitian case: $g=1.4$. 
Time evolution of the initial wave packet under the same condition as in Fig. 1.
Here,  the values of disorder strength $W$ are such that
$W=0.4, 3.6, 8.0, 10.0$, respectively, for panels (a-d).
$\theta_0=0$, $j_0=580$, and $L=601$.
}
\label{fig_1pd}
\end{figure}
It has recently been pointed out
\cite{Longhi}
that
the wave-packet spreading in non-Hermitian systems with non-reciprocal hopping
shows very different features.
In the presence of a finite non-reciprocity $g$ in hopping
the standard ``cascade-like'' wave-packet spreading 
[see Fig. \ref{fig_1herm}, panel (a)]
in the Hermitian limit $g=0$
becomes extinct.
It only revives in the vicinity of the localization transition: $W\simeq W_c$
[see Fig. \ref{fig_1pd}, panel (c)].
In the extended phase: $W<W_c$,
the wave packet does not spread,
but it only {\it slides} 
in the direction imposed by the non-reciprocality $g$.
In the body of the paper
we show and
study systematically
how this drastic change
in the manner of wave-packet spreading 
caused by the non-reciprocity $g$
is reflected in
the dynamical property of the entanglement entropy.

\begin{figure}
\includegraphics[width=85mm]{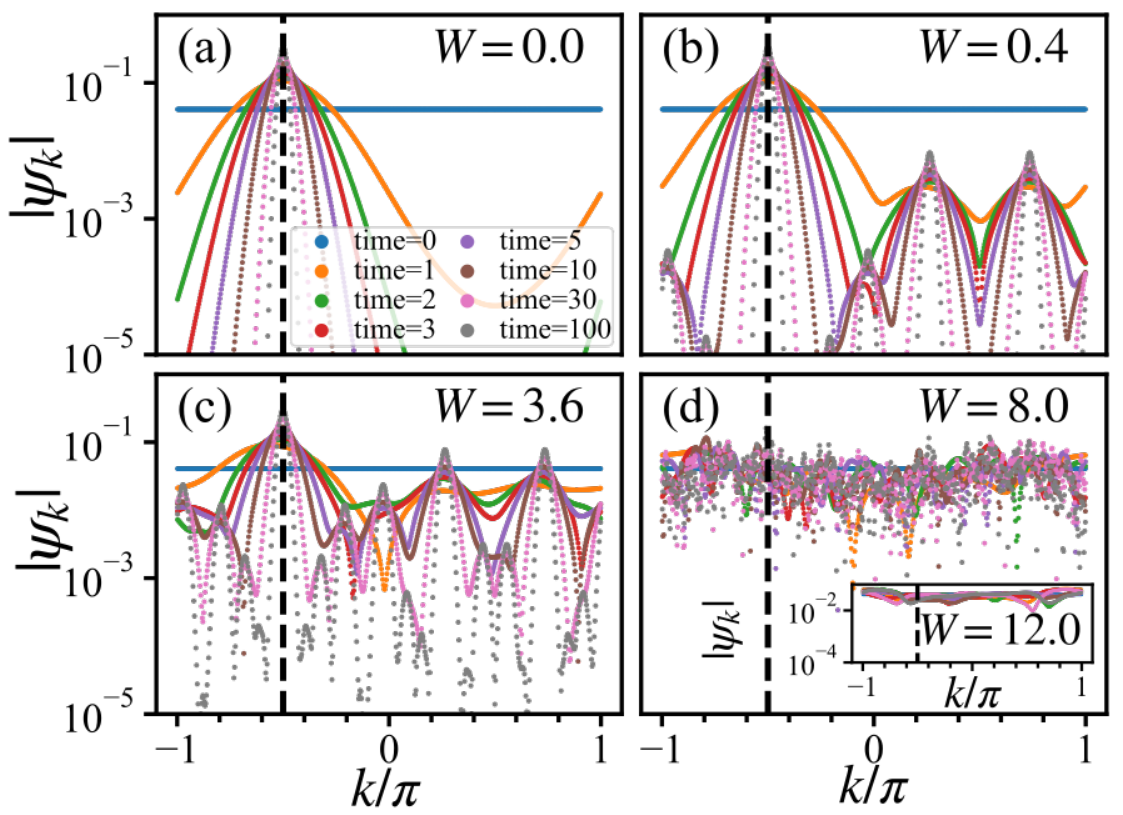}
\caption{Single-particle dynamics; 
profile of
the wave function in $k$-space.
The Fourier transform
$|\psi_k (t)|$ 
[Eq. (\ref{psi_sup})]
of the wave packet
is plotted as a function of $k$ 
at different time slices $t=0,1,2,3,5,30,100$
as indicated in the inset of panel (a).
Different panels (a-d) and the inset of panel (d) 
correspond, respectively, 
to different values of disorder strength $W=0, 0.4, 3.6, 8.0, 12.0$.
$g$ is fixed at $g=1.4$. $\theta_0=0$ (no disorder average).
Other settings 
follow those of 
Fig. \ref{fig_1pd}.
}
\label{fig_psik}
\end{figure}

As already mentioned
in the Hermitian limit,
the inter-particle interaction $V$ plays a non-trivial role
in determining the dynamical property of the entanglement entropy;
especially,
its effect is predominant
in long time-scale dynamics.
Here, we show in this paper that
the inter-particle interaction $V$ plays also
a non-trivial and principal role
in the entanglement dynamics
of a non-Hermitian system with non-reciprocal hopping.
The remainder of the paper is structured as follows.
In Sec. \ref{sectionII} we describe the model and show the mechanics of unusual dynamics.
In Sec. \ref{sectionIII} we give details of the entanglement dynamics, highlighting its non-monotonic time evolution.
In Sec. \ref{sectionIV} we examine the size dependence of the results, predicting an unusual area-volume-area law crossover of the maximal entanglement entropy.
In Sec. \ref{sectionV} we point out  and visualize the characteristic stages in the evolution of the reduced density matrix, which underlie the unusual non-Hermitian entanglement dynamics.
Sec. \ref{sectionVI} is devoted to concluding remarks.
Some details are left to the appendices.

\section{Spreading of a wave packet 
in non-Hermitian systems with non-reciprocal hopping}
\label{sectionII}

\subsection{Single-particle case: Hatano-Nelson $\times$
Aubry-Andr\'{e} model}
\begin{figure}
\includegraphics[width=85mm]{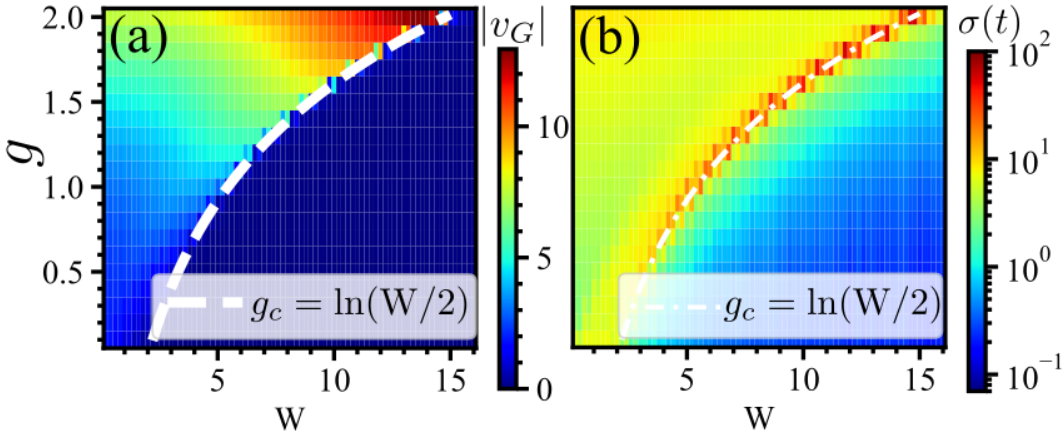}
\caption{Single-particle dynamics; velocity and density fluctuation profiles.
Panel (a) represents 
the sliding velocity $v_G=[x_G(t)-x_G(0)]/t$
[cf. Eq. (\ref{xg})]
indicated by a variation of plot colors indicated in the color bar
at a given set of parameters $g$ and $W$;
its distribution is shown in the $(W,g)$-plane.
Evaluated at $t=t_1=20$.
Panel (b) shows a similar plot for the density fluctuation, enumerated by the quantity:
$\sigma(t)=\Delta x (t)-\Delta x (0)$, 
also evaluated at $t=t_1$;
$\Delta x (t)$ is as given in Eq. (\ref{delta_x}).
The same initial state as in Fig. \ref{fig_1pd} and Fig. \ref{fig_psik}.
Average over $50$ samples
with different $\theta_0$.
}
\label{fig_1pv}
\end{figure}

Let us consider the following one-dimensional tight-binding model 
with non-reciprocal hopping amplitudes $\Gamma_L,\Gamma_R$
(Hatano-Nelson model):\cite{Hatano1}
\begin{eqnarray}
H&=&-\sum_{j=0}^{L-1}
\Big(
\Gamma_R
|j+1\rangle\langle j|
+\Gamma_L
|j\rangle\langle j+1|
\Big)
\nonumber \\
&+&W_j
\sum_{j=0}^{L-1}
|j\rangle\langle j|,
\label{ham_sp}
\end{eqnarray}
where
$\Gamma_L=e^{g} \Gamma_0$, $\Gamma_R=e^{-g} \Gamma_0$ with $g$ being
a parameter quantifying the degree of non-reciprocity.
$g$ is also
sometimes regarded as an imaginary vector potential.
\cite{Hatano1}
\footnote{
$g$ plays the same role as a damper in a Newtonian harmonic oscillator 
that tend to make the oscillation of
the wave function damp in time.
If 
$\Gamma_R=e^{-g} \Gamma_0$ ($g>0$)
corresponds to hopping in the direction of positive $x$,
the eigen wave functions tend to damp with increasing $x$,
showing a damped harmonic oscillation,
while
the regime of $g<0$ corresponds to the case of an anti-damper;
i.e., to an unrealistic parameter regime
in the Newtonian model.
}
$|j\rangle$ represents a single-particle state localized at site $j$.
In the first two (hopping) terms
we have chosen the boundary condition to be periodic:
$|L\rangle\equiv |0\rangle $.
\footnote{
In non-Hermitian quantum mechanics with non-reciprocal hopping,
choice of the boundary condition is a subtle issue, especially in its statics,
while here,
we are concerned about its dynamics
in which
the choice of the boundary condition is 
unlike in statics (see Appendix \ref{app_BC} for details)
not a central issue}
In the third term,
we have chosen the on-site potential $W_j$
to be quasi-periodic:
\footnote{
Note that
unlike the case of uncorrelated disorder; i.e., $W_j$ obeying a uniform distribution,
the addition of disorder does not lead immediately 
to localization
in the Aubry-Andre model; 
i.e., in the case of quasi-periodic disorder (\ref{W_qp})
even in the Hermitian limit: $g=0$
and
despite the one dimensionality of the model.
\cite{AALR}
This will be a merit
in the study of the entanglement entropy in the non-Hermitian case: $g\neq 0$,
since
one can always compare the situation at a finite $g\neq 0$
with the Hermitian case: $g=0$.
In the case of uncorrelated disorder,
the Hermitian limit becomes rather special.}
\begin{equation}
W_j=W \cos (2\pi\theta j+\theta_0),
\label{W_qp}
\end{equation}
playing effectively the role of a random potential (Aubry-Andr\'{e} model\cite{AAmodel}),
where
$\theta$ is an irrational constant, 
which we choose to be the so-called (inverse) golden ratio: $\theta=(\sqrt{5}-1)/2$.
$\theta_0$ is an additive phase 
introduced for the purpose of taking a disorder average;
averaging over $\theta_0$ distributed uniformly in the range $\in[0,2\pi)$ plays effectively the role of
averaging over different disorder configurations.
\footnote{In the practical calculation
$\theta_0$-averaging has been taken over $50$ samples; e.g., in the evaluation of
$x_G (t)$ and its fluctuation (Figs. \ref{fig_1pv}, \ref{fig_mpv}),
and
in the evaluation of different types entanglement entropies:
$S_{\rm tot}$, $S_{\rm num}$ and $S_{\rm conf}$.
}

In the Hermitian limit: $g=0$,
the eigenstates are extended when $W$ is weak enough ($W<W_c$),
while localized for $W>W_c$,
where 
\begin{equation}
{W_c\over2\Gamma_0}=1.
\label{wc}
\end{equation}
This may be understood 
\footnote{
The critical point (\ref{wc})
may be also understood from the viewpoint of the duality of the model.\cite{AArev}
}
from the behavior of 
localization length $\xi$ defined in the localized phase;
\cite{Longhi,rare_thermal}
i.e.,
\begin{equation}
\xi^{-1} \simeq \log {W\over2\Gamma}.
\label{xi}
\end{equation}
The localization length $\xi$ diverges as $W$ approaches the critical value (\ref{wc})
from above.

In the non-Hermitian case: $g\neq 0$,
the delocalization point is determined by the condition:\cite{Hatano2,Gil}
\footnote{
In the non-Hermitian case: $g\neq 0$, 
$g$ plays the role of an imaginary vector potential
that appears 
in the wave functions of a localized state
such that
\[
\psi^{L,R}(x)\sim \exp(-\frac{|x-x_c|}{\xi}\mp g(x-x_c)),
\]
where 
$\psi^{L,R}(x)$ 
are left and right eigenvectors;
$x_c$ is its localization center, while $\xi$ represents the corresponding localization length.
%
If $g>\xi^{-1}$,
either $\psi^L(x)$ or $\psi^R(x)$ diverges;
the wave functions
$\psi^{L,R}(x)$ 
no longer represent an exponentially localized state.
Therefore, 
the delocalized transition point is determined by the condition
(\ref{xi_g}).
}
\begin{equation}
\xi^{-1}=g>0.
\label{xi_g}
\end{equation}
If this is combined with Eq. (\ref{xi}), the delocalization transition is expected to occur
at
\begin{equation}
W=W_c=2\Gamma_0\ e^g =2\Gamma_L,
\label{wc_NH}
\end{equation}
where we have assumed $\Gamma_L>\Gamma_R$ ($g>0$);
i.e.,
$W_c$ in the non-Hermitian case
is found simply by replacing $\Gamma_0$ in Eq. (\ref{wc})
with the right/large hopping amplitude $\Gamma_0$.
Both in the Hermitian and non-Hermitian cases,
the location of the mobility edge (\ref{wc_NH})
does not depend on the energy $\epsilon_n$;
$H |n\rangle=\epsilon_n |n\rangle$.
When $g\neq 0$,
the eigenenergy $\epsilon_n$ 
becomes {\it complex} in the extended phase 
($W<W_c=2\Gamma_L$);
cf. the case of free particle motion described in Appendix \ref{app_gauss},
while it remains {\it real}
in the localized phase ($W>W_c$).
Thus, the localization-delocalization transition 
is accompanied by a real-complex transition of the eigenenergies
(see Appendix \ref{app_BC} for details).

Let us 
focus on the dynamics of the system
by following how an initially localized wave packet evolves in time.
Four panels of Figs. \ref{fig_1herm} and \ref{fig_1pd}
show examples of such dynamics.
We assume that at $t=0$
the wave packet is 
just at an initial site $j=j_0$;
\begin{equation}
|\psi (t=0)\rangle=|j_0\rangle.
\label{psi_0}
\end{equation}
At time $t$,
the wave packet may evolve as
\begin{eqnarray}
|\psi (t)\rangle&=&\sum_j \psi_j(t) |j\rangle
\nonumber \\
&=&\sum_{n} c_n e^{-i\epsilon_n t} |n\rangle,
\label{psi_t}
\end{eqnarray}
where
$|n\rangle$ represents the $n$th single-particle eigenstate of the Hamiltonian (\ref{ham_sp})
with an eigenenergy $\epsilon_n$;
i.e., $H |n\rangle=\epsilon_n |n\rangle$,
while
$c_n =\langle\langle n|\psi (t=0)\rangle$.
Here, 
$\langle\langle n|$ represents the {\it left} eigenstate corresponding to the eigenenergy $\epsilon_n$
:
$\langle\langle n|H=\epsilon_n\langle\langle n|$
and not $|n\rangle^\dagger$;
$\langle\langle n|\neq |n\rangle^\dagger$.
We make sure that the left and right eigenstates satisfy the biorthogonal condition, i.e., $\langle \langle n|m\rangle=\delta_{n,m}$.
In case of $g\neq 0$,
the eigenenergy $\epsilon_n$ is typically complex;
cf. the free particle case in Appendix \ref{app_gauss} [see also Eqs. (\ref{spec_comp}), (\ref{spec_ellip})],
so that
the time-evolved wave packet 
$|\psi (t)\rangle$
literally as given in Eq. (\ref{psi_t})
tends to grow 
exponentially; its norm $\langle \psi (t)|\psi (t)\rangle$ is not conserved
due to the contribution from states with Im $\epsilon_n >0$. 
In the actual computation, 
we, therefore, rescale (renormalize) $|\psi (t)\rangle$
at every interval $\Delta t$
as
\footnote{In the actual computation,
$\Delta t$ has been chosen as
$\Delta t\simeq 10^{-4}-10^{-1}$;
id. in the multi-particle case.}
\begin{equation}
|\psi (t)\rangle\rightarrow|\tilde{\psi}(t)\rangle=
{|\psi (t)\rangle\over\sqrt{\langle\psi (t)|\psi (t)\rangle}}.
\label{renorm}
\end{equation}
and avoid this computational difficulty.\cite{Hamaz,china}

Let us first consider the Hermitian case. 
The four panels of
Fig. \ref{fig_1herm} show the distribution of $|\psi_j(t)|$ in the case of Hermite case for different strength of $W$.
At site $j$ (the abscissa) and at time $t$ (the ordinate), the amplitude of $|\psi_j(t)|$ is specified
by a variation of the plot color indicated in the color bar.
In the 
clean limit [Fig. \ref{fig_1herm}(a)], 
the wave packet spreads symmetrically in the two directions.
As the strength $W$ of disorder increases
(Fig. \ref{fig_1herm}(b), (c)), 
spreading of the wave packet tends to become slower. 
Finally, beyond the critical disorder strength (Fig. \ref{fig_1herm}(d)), 
the wave packet ceases to spread. 
Then,
the four panels of
Fig. \ref{fig_1pd} show the distribution of $|\psi_j(t)|$ in the case of 
$g=1.4$
($L=601$, $j_0=580$)
for different strength of $W$.
Unlike in the Hermitian case
\footnote{
In the Hermitian limit ($g=0$)
the wave packet {\it spreads}
as time progresses, unless
disorder $W$ is not too strong ($W/J<2$); see e.g., Fig. 2 of Ref. \onlinecite{Longhi}.
In a non-Hermitian system ($g\neq 0$)
wave packet spreading practically ceases,\cite{Longhi}
at least
the one as in the Hermitian limit;
i.e., a cascade-like spread of the wave packet.
In any case, the dynamics
becomes very different from the
Hermitian case.
Here, we further clarify this point.
}
(Fig. \ref{fig_1herm})
the four panels show
that
the wave packet does not spread;
at least in the regime of weak $W$ [cases of panels (a-b)],
\footnote{at least in the short time scale;
in the (long) time regime
at which the imaginary part Im $\epsilon_n$ comes into play,
$|\psi (t)\rangle$ decays into a single eigenstate with a maximal Im $\epsilon_n$;
see Appendix \ref{app_gauss} for details.
In such a long time scale the wave packet 
may spread in time, but not in the sense considered here.
}
but rather {\it slides} 
in the direction imposed by the non-reciprocity $g$. 
%
In the non-Hermitian case $g\neq 0$
the standard cascade-like wave packet spreading
as in the Hermitian limit disappears
in the regime of weak $W$ [panels (a-b)]
such that $W\ll W_c$,\cite{Longhi}
but a similar (cascade-like) behavior reappears
in the vicinity of the localization transition: $W\simeq W_c$
[panel (c)].
In the localized phase, the wave packet does not move
[panel (d)].
Comparing the three cases on the delocalized side
[panels (a-c)],
one also notices that the ``sliding velocity'' of the wave packet;
at least the velocity of the wave front $v_f$,
tends to increase as $W$ is increased.\cite{Longhi}

To understand 
why in the non-Hermitian system
the wave-packet dynamics
becomes very different from the standard Hermitian case,
one may well start with the clean limit: $W=0$.
In this limit,
the eigenstates are plane waves $\langle j|k\rangle=e^{ikj}/\sqrt{L}$
so that
\begin{eqnarray}
|\psi(t)\rangle 
&=& \sum_k e^{-i\epsilon_k t}|k\rangle \langle k|j_0\rangle 
\left(\equiv\sum_k \psi_k(t)|k\rangle\right)
\nonumber\\
&=& \frac{1}{\sqrt{L}} \sum_{j}\sum_{k} e^{-i\epsilon_k t+ik(j_0-j)}|j\rangle,
\label{psi_sup}
\end{eqnarray}
i.e.,
$|\psi(t)\rangle$ is generally expressed as a superposition of
such plane waves;
at site $j$
contributions from different $k$ add up
with a phase factor
\begin{equation}
e^{i\phi (k)}=e^{-i\epsilon_k t+ik(j_0-j)}.
\end{equation}
At $t=0$ and at $j\neq j_0$,
such contributions are out of phase and cancel each other,
while
at $j=j_0$ they add up in phase to form the peak of the initial wave packet.
At $t>0$, similarly,
the only non-vanishing contributions [in the summation over $k$ in Eq. (\ref{psi_sup})]
are those from the neighborhood of $k=\overline{k}$
at which the phase $\phi(k)$ becomes {\it stationary}; i.e.,
$\phi'(\overline{k})=0$, or
\begin{equation}
2\Gamma_0\sin \overline{k}\ t = j-j_0.
\label{stat}
\end{equation}
Since $|\sin \overline{k}| \le 1$, $|j-j_0|=2\Gamma_0 t \equiv v_f t$ 
defines the position of the wave front, or a ``light cone''.
\cite{Longhi}
In the Hermitian limit
the initially localized wave packet spreads linearly in time:
$\Delta x (t) \propto t$; i.e.,
$\Delta x (t) \sim t^\sigma$ with the exponent $\sigma\simeq 1$,
\footnote{
Note that in the classical diffusion dynamics
this exponent $\sigma$ is 1/2
in one spatial dimension,
showing a specific square-root scaling.
In a diffusive metal the electron density in a random environment 
obeys effectively the same dynamics, 
leading to diffusive conduction in a metal. 
Here, in our setup we focus on a coherent 
quantum dynamics of a wave packet.
In this case, clearly, neither the
the wave function itself $\psi(x,t)$ nor
the density profile $|\psi(x,t)|^2$
obeys a diffusion equation.
}
where
\begin{equation}
\Delta x (t)=\sqrt{\sum_j (j-j_0)^2 |\psi_j(t)|^2}.
\label{delta_x_0}
\end{equation}
represents
the spread of the light cone.
Addition of disorder $W$ suppresses the wave-packet spreading;
as shown in Fig. 2 of Ref. \onlinecite{Longhi},
the velocity 
$v= \Delta x (t) /t$,
characterizing the speed of the spreading of wave packet
is shown to decrease linearly with $W$ 
and vanishes at $W=W_c$.

On addition of non-Hermiticity $g\neq 0$,
a different mechanism or a principle
sets in to play a role
in the wave-packet dynamics of Eq. (\ref{psi_sup}),
since
the eigenenergies $\epsilon_k$ become {\it complex}:
\begin{equation}
\epsilon_k=-2\Gamma_0\cos(k-ig),
\label{spec_comp}
\end{equation}
which
in the complex energy plane,
take values on an ellipse:
\begin{equation}
\Big({{\rm Re}\ \epsilon_k\over \Gamma_0\cosh g}\Big)^2+
\Big({{\rm Im}\ \epsilon_k\over \Gamma_0\sinh g}\Big)^2=1.
\label{spec_ellip}
\end{equation}
In this case
contributions from those $k$'s which have
maximal Im $\epsilon_k$'s
become more important in the superposition (\ref{psi_sup}).
In case of Eq. (\ref{spec_comp}) [cf. also Eq. (\ref{spec_ellip})] 
such $k$'s are found at (around)
\begin{equation}
k=k_0=-\pi/2.
\end{equation}
Thus,
in the non-Hermitian (free-particle) dynamics,
the initial state (\ref{psi_0}) {\it dissolves}
in the course of time evolution (\ref{psi_sup})
into a Gaussian wave packet:
\begin{eqnarray}
|\psi(t)\rangle
&\simeq&
\sum_{j}|j\rangle \exp(-\frac{((j_0-j)+2(\cosh g)t)^2}{4(\sinh g)t})
\nonumber\\
&&\times e^{2(\sinh g)t}/\sqrt{4(\sinh g)t},
\label{psi_gauss}
\end{eqnarray}
which
are composed of plane waves with $k$'s
found around
$k=k_0$;
as for the derivation of Eq. (\ref{psi_gauss_app}), see Eq. (\ref{psi_gauss}) and related arguments in
Appendix \ref{app_gauss}.
Remarkably,
the resulting Eq. (\ref{psi_gauss})
is a wave packet that
{\it slides} in the direction imposed by $g$,
though its expanse
gradually increases as time evolves 
Since
as a guiding principle,
\begin{enumerate}
\item 
the survival of Max Im $\epsilon_k$ 
has priority over 
\item
the stationary phase condition 
[cf. Eq. (\ref{stat}) in the Hermitian case],
\end{enumerate}
the non-Hermitian (free-particle) dynamics
is fully governed by the principle 1.,
unlike in the Hermitian case
in which the principle 2. becomes manifest 
under the condition that
the principle 1. is disabled 
and masked.
In Fig. \ref{fig_psik}, panel (a)
the distribution of $|\psi_k(t)|$ at some fixed $t$'s are shown.
$|\psi_k(t)|$ shows a Gaussian type distribution centered at $k=k_0$,
and its width tends to become narrower 
as $t$ evolves 
[cf. Appendix \ref{app_gauss}].

Panels (b-e) of Fig. \ref{fig_psik} show
how the addition of disorder $W$
affects and eventually destroys
this peak structure of $\psi_k$.
In panel (b)
two side peaks may be conspicuous at $k=k_1, k_2$;
they are associated with 
the quasi-periodic nature of the potential (\ref{W_qp});
Bloch waves of these $k$'s are quasi-commensurate with the potential.
The complex energy spectrum $\epsilon_k$
also shows 
(in the presence of $W\neq 0$) 
[cf. Eqs. (\ref{spec_comp}), (\ref{spec_ellip}) in the potential-free case ($W=0$)]
an extremum at theses $k$'s,
showing local maxima of Im $\epsilon_k$.\cite{Longhi}
As $W$ is increased, such side peaks multiply [panel (c)], 
\footnote{Let us emphasize, here, that
the side peaks ($k=k_1, k_2, \cdots$)
are due to the quasi-periodic potential (\ref{W_qp}),
while
the main peak  ($k=k_0$)
stems from
the survival of Max Im $\epsilon_k$,
and the two types of peaks have a different nature.
}
and 
the system gradually evolves into
the cascade regime represented by panel (d),
where
the distribution of $|\psi_k(t)|$ is almost uniform, but still
there are plenty of 
tiny peaks,
while
in the localized regime [in panel (e)]
the distribution becomes flat and smooth.

To further quantify features 
specific to the non-Hermitian wave packet dynamics,
it may be natural to focus on
\begin{enumerate}
\item 
how fast
the center of the gravity 
\begin{equation}
x_G(t)=\sum_j j |\psi_j(t)|^2
\label{xg}
\end{equation}
of the wave packet moves,
and also
\item
to what extent the wave packet is spread around $x_G(t)$.
\end{enumerate}
It turns out that
$x_G(t)\simeq x_G(0)+v_G t$
so that 1. can be measured by the velocity $v_G$.
In panel (a) of Fig. \ref{fig_1pv}, 
the magnitude of $v_G$ is
plotted
(determined by evaluating $x_G(t)$ at $t=t_1$
\footnote{
The measurement time $t_1$
has been chosen to be a value ($t=t_1=20$ in Fig. \ref{fig_1pv})
such that
the wave packet does not travel across the (periodic) boundary.
}
)
as changing the set of parameters $(W,g)$,
and is indicated by a variation of plot color.
The plot shows that
$v_G$ is finite in the extended phase $W<W_c$, 
while it practically vanishes in the localized phase $W>W_c$.
The location of the phase boundary (\ref{wc_NH}),
or equivalently,
\begin{equation}
g=\log {W\over 2}
\label{wc_NH2}
\end{equation}
is indicated by a broken curve in the panel.
On the side of the extended phase $W<W_c$, 
$v_G$
continues to take a relatively large value
until quite close to the phase transition;
at a fixed value of $g$,
it rather tends to increase 
as $W$ increases
until an abrupt fall at the phase transition.

Panel (b) shows a similar plot for the quantity: 
$\sigma(t)=\Delta x (t)-\Delta x (0)$ at some fixed time $t=t_0$,
where $\Delta x (t)$ has been redefined as
\begin{equation}
\Delta x (t)=\sqrt{\sum_j (j-x_G(t))^2|\psi_j (t)|^2}.
\label{delta_x}
\end{equation}
The quantity $\sigma(t)=\Delta x (t)-\Delta x (0)$ is expected to measure
2. to what extent the density $\rho_j(t)$ is spread around $x_G(t)$
The plot shows that 
similarly to the behavior of $v_G$ in panel (a),
the spread of the wave packet
also shows a sharp distinction
in the extended ($W<W_c$) and localized ($W>W_c$) phase.
Here,
in terms of $\Delta x (t)$, 
not only
it takes a finite value on the side of the extended phase: $W<W_c$, 
but 
the appearance of a peak may be easily seen
as $W$ is increased toward and close to the localization transition (\ref{wc_NH2})
at $g$ fixed.
We interpret that
this enhancement of $\sigma(t)$ slightly before the localization transition
reflects the
the cascade-like explosion of the wave packet
seen
in the density profile in Fig. \ref{fig_1pd}, panel (c).

\begin{figure}
\includegraphics[width=80mm]{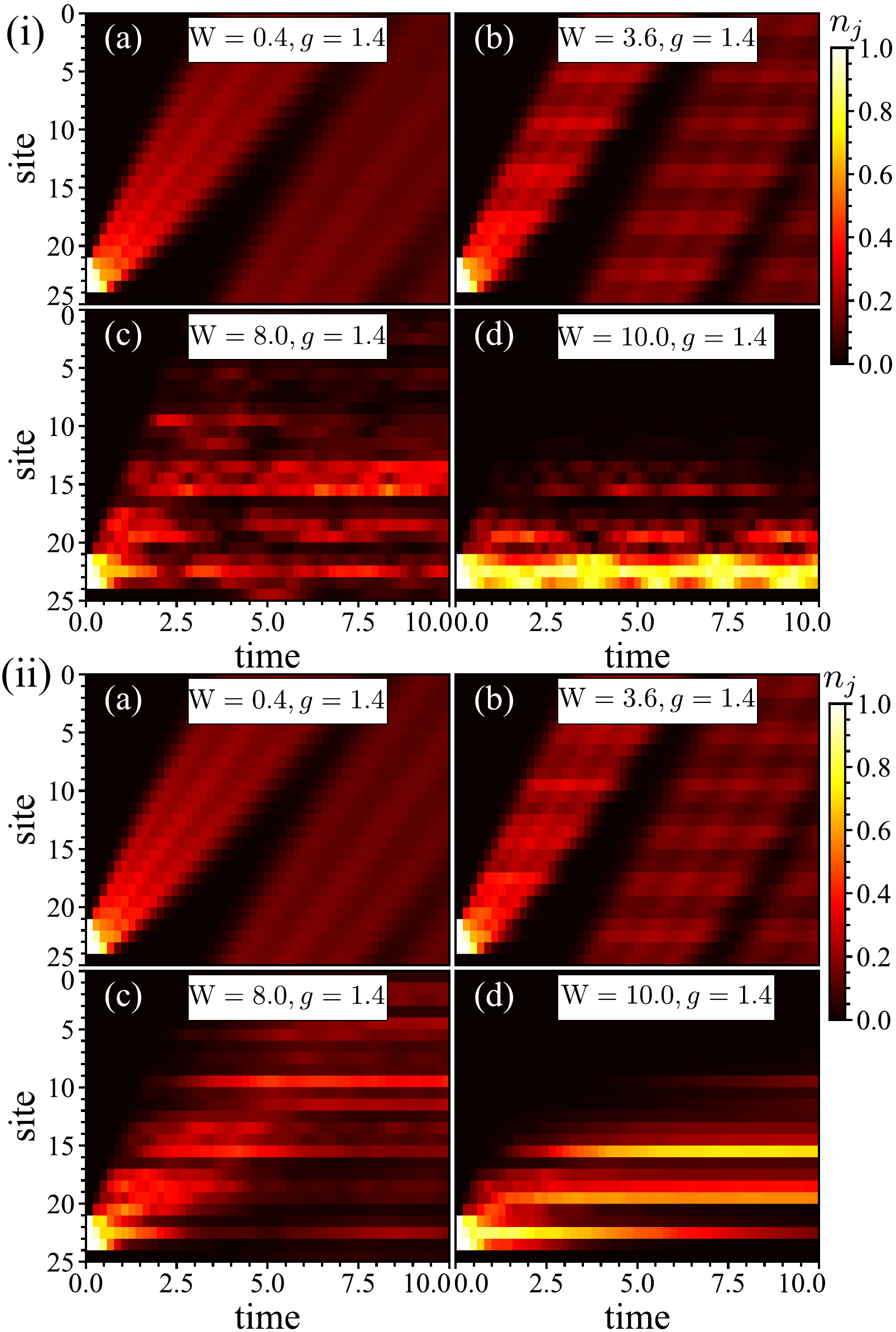}
\caption{Multi-particle dynamics; 
evolution of the initial wave packet chosen to be in a domain wall form
Eq. (\ref{d_wall}).
The eight panels of Fig. \ref{fig_mpd} show the evolution of
the particle density
$n_j(t)$ as given in Eq. (\ref{njt})
at site $j$ and at time $t=t_1$; here $t_1$ is chosen as $t_1=2.2$,
by a gradation of plot colors indicated in the color bar.
The system size $L$ is set to $L=25$.
Different panels (a-d) correspond to different values of disorder strength $W$;
$W=0.4, 3.6, 8.0, 10.0$, respectively, for panels (a-d).
$g$ is fixed at $g=1.4$. $\theta_0=0$ (no disorder average).
The four upper panel in the case (i)
represent the non-interacting case: $V=0$, while
those in (ii)
represent an interacting case: $V=2$.
}
\label{fig_mpd}
\end{figure}

\begin{figure}
\includegraphics[width=85mm]{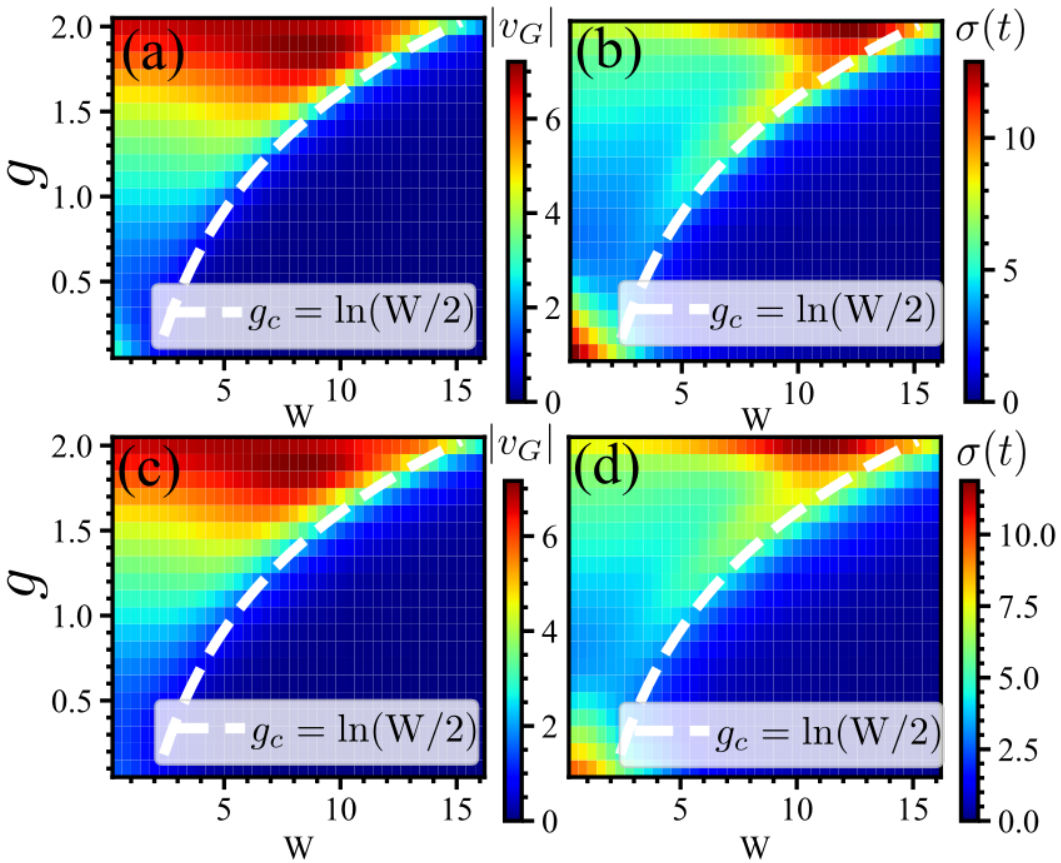}
\caption{Multi-particle dynamics; velocity and density fluctuation profiles.
Panels (a), (c) show 
distribution of
the sliding velocity $v_G$,
in the parameter space $(W,g)$,
while
panels (b), (d) represent a similar plot
for 
$\sigma(t)=\Delta x (t)-\Delta x (0)$.
$v_G$ and $\Delta x (t)$ are
evaluated as 
in Eqs. (\ref{xg_mp})
in the multi-particle case.
Panels (a-b) represents the non-interacting case: $V=0$,
panels (c-d) an interacting case: $V=2$.
The same initial state as in Fig. \ref{fig_mpd};
average over $50$ samples with different $\theta_0$.
}
\label{fig_mpv}
\end{figure}

\subsection{Case of an interacting system}

Here, we consider whether or not/how the presence of interaction 
may affect the above non-interacting picture.
As a concrete model,
we have employed
the following bosonic version of the
Hatano-Nelson $\times$ Aubry-Andr\'{e} model with a nearest neighbor inter-particle interaction $V$:
\begin{eqnarray}
{\cal H}&=& - \sum_{j=0}^{L-1}
\Big(\Gamma_L b_j^\dagger b_{j+1}+\Gamma_R b_{j+1}^\dagger b_{j}\Big)
\nonumber \\
&+&\sum_{j=0}^{L-1}
\Big(V \hat{n}_j \hat{n}_{j+1}+W_j \hat{n}_j\Big),
\label{ham_mp}
\end{eqnarray}
where $b_j^\dagger$ ($b_j$) creates (annihilates) a particle 
at site $j$, 
while $\hat{n}_j=b_j^{\dagger}b_j$ counts
the number $n_j$ of such particles
found at site $j$.
Following Refs. \onlinecite{Hamaz}, \onlinecite{china},
we assume that
our particles are hard-core bosons: $n_j=0,1$.
\footnote{
In case they are true fermions,
anticommutation of $b_j$ and $b_j^\dagger$
may lead to a quantitatively different result
under a periodic boundary condition, especially in the delocalized regime;
see Appendix \ref{app_fermi} for details.
}

Fig. \ref{fig_mpd} shows the examples of multi-particle dynamics in this system
when the initial state is chosen to be the following {\it domain wall} state
\begin{equation}
|\Psi (t=0)\rangle = |00\cdots 0 11\cdots 1\rangle,
\label{d_wall}
\end{equation}
i.e., the last $N_b$ sites are occupied in the local basis;
\footnote{
In the actual simulation, we have shifted 
the domain of occupied sites; i.e., $j$'s such that $n_j=1$
slightly (actually, by one site) to the inner side of the system
so that the wave function does not spread across the boundary.}
$N_b$ represents the number of particles (bosons)
in the system; 
in this numerical shown in Fig. \ref{fig_mpd}
it is chosen as $N_b=3$.
%
At time $t$, the initial state (\ref{d_wall}) evolves as
\begin{equation}
|\Psi (t)\rangle=\sum_{\mu} c_\mu e^{-iE_\mu t} |\mu\rangle,
\label{Psi_sup}
\end{equation}
where
$|\mu\rangle$ is the eigenstate of the Hamiltonian Eq. (\ref{ham_mp});

$E_\mu$ is the corresponding eigenenergy:
${\cal H}|\mu\rangle=E_\mu |\mu\rangle$,
while
$c_\mu =\langle\langle\mu|\Psi (t=0)\rangle$.
Here, 
$\langle\langle\mu|$ represents the {\it left} eigenstate corresponding to the eigenenergy $E_\mu$: 
$\langle\langle\mu |{\cal H}=E_\mu\langle\langle\mu|$
and not $|\mu\rangle^\dagger$;
$\langle\langle\mu |\neq |\mu\rangle^\dagger$.
We make sure that the left and right eigenstates satisfy the biorthogonal condition, i.e., $\langle \langle \mu|\nu\rangle=\delta_{\mu,\nu}$.
These are analogous to the single-particle case.
Also, as the single-particle eigenenergy $\epsilon_n$ is complex in general,
so is the many-body eigenenergy $E_\mu$.
As a result, the time-evolved wave packet $|\Psi (t)\rangle$
as given in Eq. (\ref{Psi_sup})
tends to grow exponentially.
To avoid this computational difficulty, 
we rescale (renormalize) $|\Psi (t)\rangle$
in the same way as in Eq. (\ref{renorm})
at every interval $\Delta t\simeq 10^{-4}-10^{-1}$
in the actual computation.
The eight panels of Fig. \ref{fig_mpd} show the evolution of
the particle density
\begin{equation}
n_j(t)=\langle \Psi (t) |\hat{n}_j |\Psi (t)\rangle
\label{njt}
\end{equation}
at site $j$ at time $t$
by a color variation; 
the higher is the density, the brighter the color is.
The system size $L$ is set to $L=25$.
The four panels in the upper case (i)
represents the non-interacting case: $V=0$, while
those in the lower case (ii)
represents an interacting case: $V=2$.
In these panels one can still see the tendency observed
in the single-particle dynamics;
e.g., the cascade-like feature in wave packet spreading
can be seen in panels (c) [both in (i) and (ii)]. 

Figs. \ref{fig_mpv}
represent
the distribution of
$v_G$ and $\sigma(t)=\Delta x (t)-\Delta x (0)$
in the parameter space: $(W,g)$;
the two figures correspond, respectively, to the cases of $V=0$ and $V=2$.
$v_G$ and $\sigma(t)$
are calculated similarly in the single-particle case;
i.e., via, 
\begin{eqnarray}
x_G(t)&=&\sum_j j n_j(t) \simeq x_G(0)+v_G t,
\nonumber \\
\Delta x (t)&=&\sqrt{\sum_j (j-x_G(t))^2 n_j (t)/\sum_j n_j (t)}.
\label{xg_mp}
\end{eqnarray}
In both of the figures
the two quantities $v_G$ and $\sigma(t)$,
both take a finite value on the side of the extended phase: $W<W_c$,
while they vanish on the localized side: $W>W_c$.
In $\sigma(t)$ [panels (b)],
one can also clearly see an enhancement before the localization transition.
Thus,
the specific and unusual features in wave packet spreading 
characteristic to non-Hermitian systems
found in the single-particle dynamics
are, though somewhat masked by the inter-particle interaction $V$,
essentially maintained
in the multi-particle dynamics.

\begin{figure*}
\includegraphics[width=170mm]{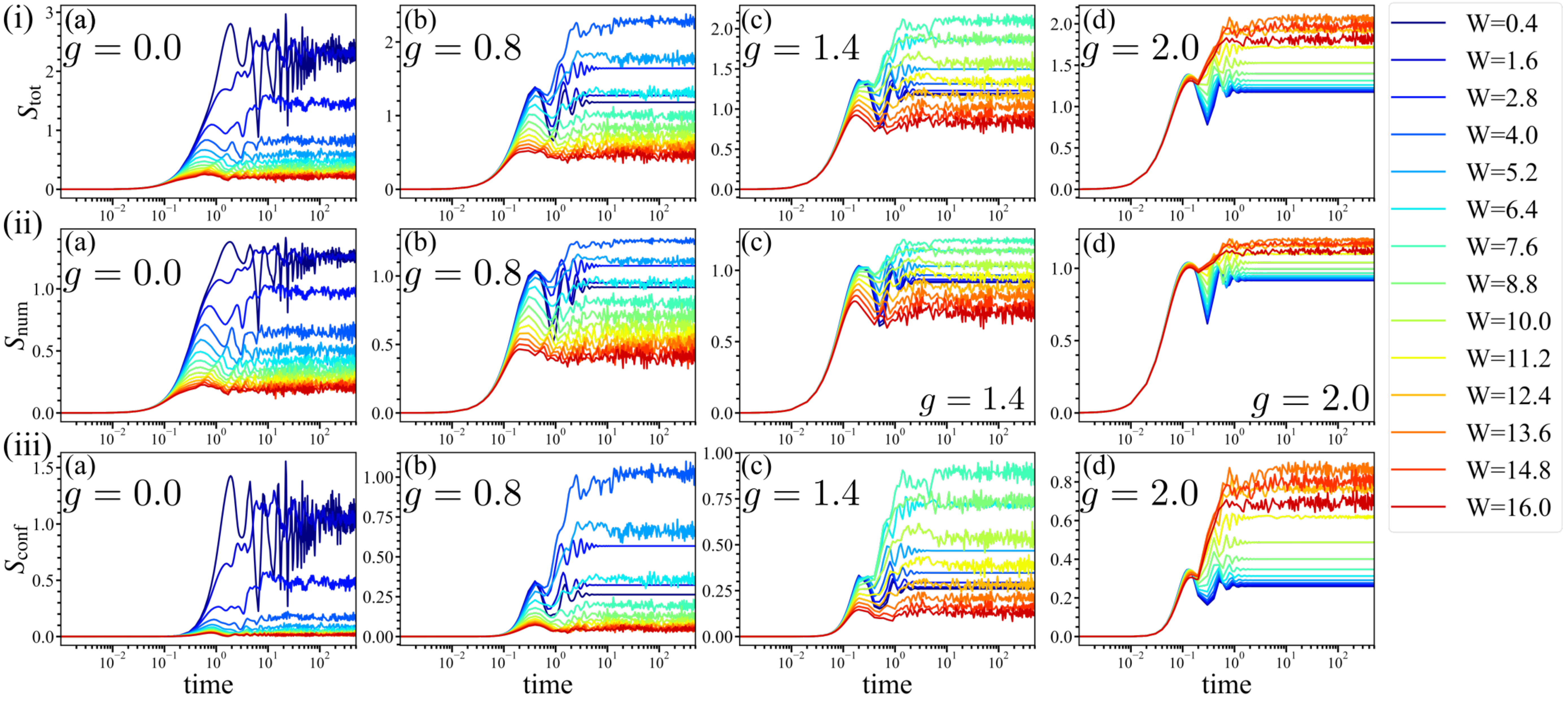}
\caption{
Entanglement dynamics; non-interacting case ($V=0$).
Three panels (i-a), (ii-a), (iii-a) in the first column 
represent the Hermitian case $g=0$, while
other panels represent non-Hermitian cases $g\neq 0$;
$g=0.8$ for panels (i-b), (ii-b), (iii-b) in the second column (in this case, $W_c=2\Gamma_L=2\Gamma_0 e^g \simeq 4.5$),
$g=1.4$ for panels (i-c), (ii-c), (ii-d) in the third column ($W_c\simeq 8.1$),
$g=2$ for panels (i-d), (ii-d), (iii-d) in the fourth column ($W_c\simeq 14.8$).
Four panels (i-a), (i-b), (i-c), (i-d) in the first raw
represent the evolution of the total entanglement entropy $S_{\rm tot}(t)$,
while
panels (ii-a), (ii-b), (ii-c), (ii-d) in the second raw, and
panels (iii-a), (iii-b), (iii-c), (iii-d) in the third raw
represent, respectively,
the number and configuration entropies, 
$S_{\rm num}(t)$
and
$S_{\rm conf}(t)$.
The system size is $L=12$.
Density-wave type initial state (\ref{d_wave}); average over $50$ samples
with different $\theta_0$.
}
\label{fig_EE_V0}
\end{figure*}

\begin{figure}
\includegraphics[width=85mm]{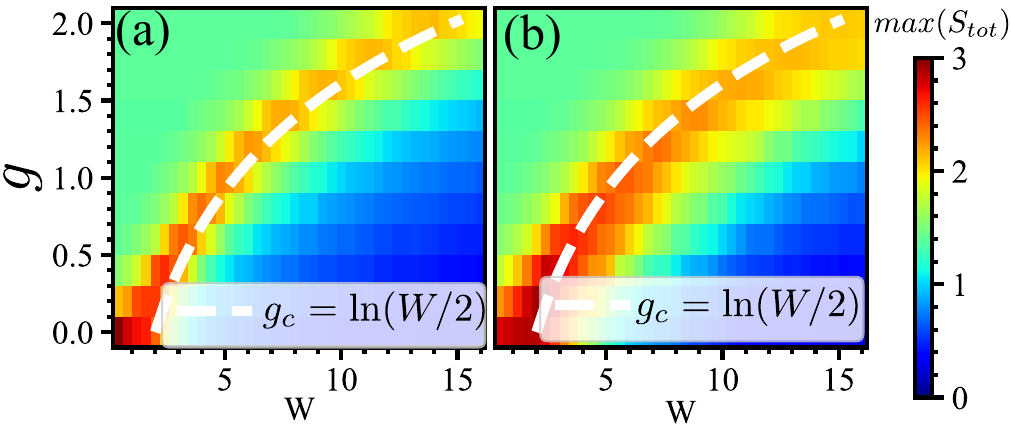}
\caption{
Distribution of the maximal entanglement entropy, Max $S_{\rm tot}(t)$
in the $(W,g)$ parameter space.
The maximal value of $S_{\rm tot}(t)$
in the time evolution is plotted
(a) non-interacting case: $V=0$ 
(the corresponding time evolution is plotted in Fig. \ref{fig_EE_V0}),
(b) interacting case: $V=2$
(id. in Fig. \ref{fig_EE_V0}).
The location of
the delocalization-localization transition in the non-interacting case:
$g=\log W/2$ [as given in Eq. (\ref{wc_NH2})]
is indicated by a broken curve (in white)
as a guide for the eyes.
Based on the same data as in Figs. \ref{fig_EE_V0}, \ref{fig_EE_V}.}
\label{fig_Smax}
\end{figure}

\begin{figure}
\includegraphics[width=50mm]{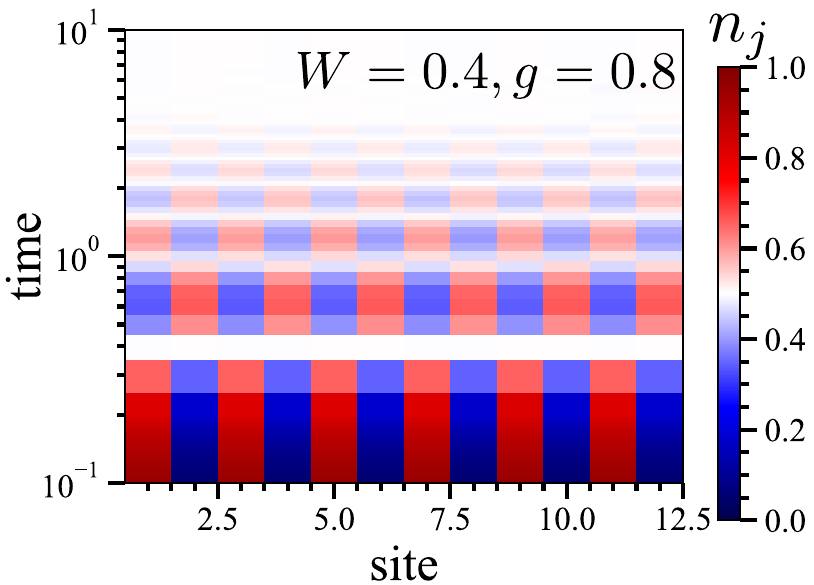}
\caption{
Damped oscillatory behavior in the density profile $n_j(t)$.
Evolution of the particle density $n_j(t)=\langle \Psi (t) |\hat{n}_j |\Psi (t)\rangle$
as given in Eq. (\ref{njt})
is plotted for the density-wave type initial state Eq. (\ref{d_wave}).
The magnitude of $n_j(t)$ at site $j$ (abscissa) and at time $t$ (ordinate) is expressed
by a variation of plot color indicated in the color bar.
g=0.8$, $W=0.4$, \theta_0=0$, $L=12$; 
no disorder averaging.
}
\label{fig_damp}
\end{figure}

\begin{figure*}
\includegraphics[width=170mm]{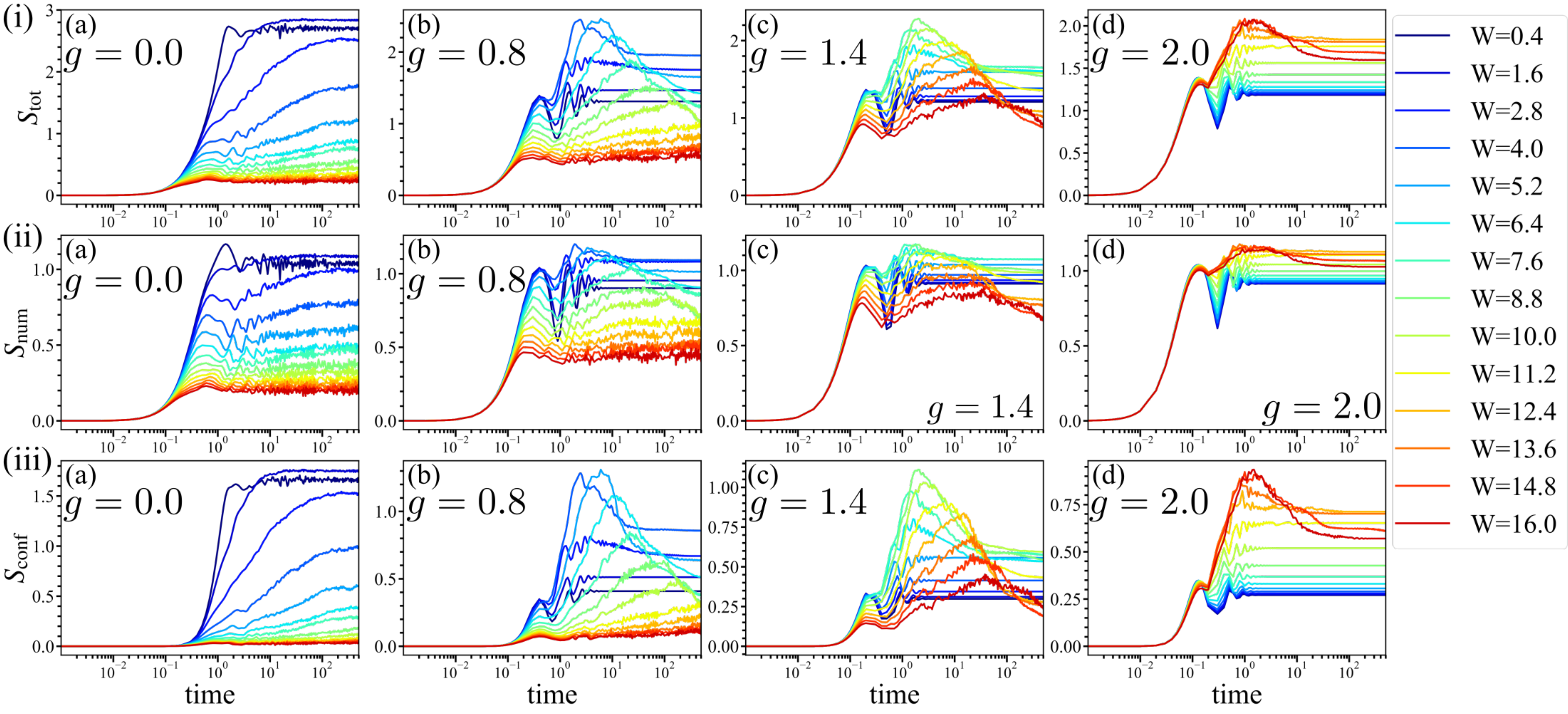}
\caption{
Entanglement dynamics; interacting case ($V=2$).
Similarly to the case of Fig. \ref{fig_EE_V0}, 
three panels (i-a), (ii-a), (iii-a) in the first column 
represent the Hermitian case $g=0$, while
other panels represent non-Hermitian cases $g\neq 0$;
$g=0.8$ for panels (i-b), (ii-b), (iii-b) in the second column,
$g=1.4$ for panels (i-c), (ii-c), (iii-c) in the third column,
$g=2$ for panels (i-d), (ii-d), (iii-d) in the fourth column.
The precise value of 
$W_c$ is unknown in the interacting case ($V\neq 0$),
but will be not far from (slightly larger than) the non-interacting values given 
in the main text and in the caption of Fig. \ref{fig_EE_V0}.
%
$L=12$.
Other conditions are the same as in Fig. \ref{fig_EE_V0}; average over $50$ samples
with different $\theta_0$.
}
\label{fig_EE_V}
\end{figure*}

\section{The entanglement dynamics}
\label{sectionIII}

The entanglement entropy
measures how quasiparticles spread and entangle
a quantum states in the system\cite{Calabrese1,Calabrese2,Calabrese3}.
In a non-interacting system,
this signifies how a wave function spreads in the system; 
the entanglement entropy increases
as the wave function spreads in the system.
In dynamics, this occurs in the process of wave packet spreading.
The initial growth of the number entropy $S_{\rm num}(t)$
in the quench dynamics
is essentially due to this effect.
In Hermitian systems with reciprocal hopping $\Gamma$
an initially localized wave packet 
spreads symmetrically in the two directions
in the clean limit
and
after a duration $t$
the wave function becomes extended to a region
of width $\Delta x \sim 2 v_f t$,
where
$v_f$ is the velocity of the wave front: $v_f\simeq J$;
\cite{Longhi}
the entanglement entropy also increases as
$S(t)\propto\Delta x \sim 2 v_f t$.
In interacting systems,
the same argument applies to quasiparticles,
with being the notion of
$v_f$ replaced with the Lieb-Robinson velocity.\cite{Lieb}
The interaction also makes the superposition of eigenstates
``nontrivial'' in the time evolution of  a wave packet,
leading to generation of entanglement (entropy).  
In the presence of disorder $W$,
\footnote{
Here, we have in mind the case of the Aubry-Andre model [cf. Eq. (\ref{W_qp})]
in which
addition of disorder does not lead immediately 
to localization
in spite of the one dimensionality of the system.
\cite{AALR}
}
as its strength is increased,
wave packet spreading tends to be suppressed;
$v_f$ tends to be suppressed
and at some critical value $W=W_c$,
$v_f$ vanishes.
When $W>W_c$,
the wave function is localized 
and wave packet spreading is essentially suppressed.
As for the behavior of $S_{\rm tot}(t)$,
it experiences a rapid growth
in the time scale wave packet spreading prevails in the system
in the extended phase,
while
such a growth is much suppressed in the localized phase [see Fig. \ref{fig_EE_V0}, panel (i-a)].
\cite{Abanin_log}


Having in mind how wave packet spreads
in non-Hermitian systems with non-reciprocal hopping,
we proceed to the analysis of
how
the entanglement entropy $S_{\rm tot}(t)$ 
evolves in time
in such systems.
We employ the same many-body Hamiltonian (\ref{ham_mp})
as in the analysis of wave packet spreading,
%
while the initial (many-body) wave packet
is chosen to be the following {\it density wave} form
(or N\'eel form in the pseudo-spin language
\cite{panda}):
\begin{equation}
|\Psi (t=0)\rangle = |101010\cdots\rangle,
\label{d_wave}
\end{equation}
where on the right hand side of the equation
we have employed the computational basis $|\nu\rangle=|n_1 n_2 \cdots n_L\rangle$;
$n_j=0,1$ represents occupation of the $j$th site.
\footnote{
We assume that our particles are hard-core bosons unless otherwise mentioned;
cf. Appendix \ref{app_fermi}.
}

The entanglement entropy $S(t)$ 
can be found by tracing out ``a half'' of the system;
i.e., 
we divide the system of length $L$ into 
two subsystems $A$ and $B$ of length $L/2$,
where
a site $j$ in $A$ satisfies $j\in j_A=\{1,2,\cdots, L/2\}$, 
while $j$ in $B$ satisfies $j\in j_B=\{L/2+1,L/2+2,\cdots, L\}$.
We then,
by tracing out the subsystem $B$,
calculate the entanglement entropy $S_A (t)$ for the subsystem $A$.
To concretize this procedure, we employ the density matrix,
\begin{equation}
\Omega (t) = |\Psi (t)\rangle \langle\Psi (t)|,
\label{DM}
\end{equation}
and perform its partial trace:
\begin{equation}
\Omega_A (t)=\rm{Tr}_B\ \Omega(t),
\label{RDM}
\end{equation}
where
in Eq. (\ref{DM})
$|\Psi (t)\rangle$ represents a many-body state at time $t$
evolved from Eq. (\ref{d_wave}).
Then,
we introduce the entanglement entropy: 
\begin{equation}
S_A(t) = - {\rm Tr}\ \Omega_A (t)\log\Omega_A (t),
\end{equation}
which is
related to the eigenvalues $\lambda_\alpha(t)$
of the reduced density matrix $\Omega_A (t)$ as
\begin{equation}
S_A(t) = - \sum_\alpha \lambda_\alpha(t) \log \lambda_\alpha(t).
\label{S_A}
\end{equation}



\subsection{Two types of contributions to the entanglement entropy:
number and configuration entropies}
\label{S_numconf}

If the number of particles in subsystem $A$, i.e., 
\begin{equation}
N_A=\sum_{j\in j_A} n_j
\end{equation}
is a good quantum number in the subsystem;
i.e., if
\begin{equation}
[\Omega_A,N_A]=0
\label{commu}
\end{equation}
holds,
the contributions to the entanglement entropy 
$S_{\rm tot}(t)=S_A(t)$
can be divided into two parts;
they are the number and configuration entropies:
\cite{science}
\begin{equation}
S_{\rm tot}(t)=S_{\rm num}(t)+S_{\rm conf}(t),
\end{equation}
where
\begin{eqnarray}
S_{\rm num}&=&-\sum_{N_A}p_{N_A} \log\ p_{N_A},
\label{S_num}
\\
S_{\rm conf}&=&-\sum_{N_A}\sum_{\alpha}
p_{N_A}\tilde{\lambda}^{(N_A)}_{\alpha}\log\ \tilde{\lambda}^{(N_A)}_{\alpha}.
\label{S_conf}
\end{eqnarray}
To find the formulas (\ref{S_num}), (\ref{S_conf}),
let us first notice that
the reduced density matrix $\Omega_A$
is block diagonal
since
$N_A$ and $\Omega_A$ are simultaneously diagonalizable [cf. Eq. (\ref{commu})];
i.e.,
\begin{equation}
\Omega_A=\Omega_{N_1}\oplus\Omega_{N_2}\oplus\Omega_{N_3}\oplus\cdots
\label{RDM_bd}
\end{equation}
where
$\Omega_{N_A}$'s are 
diagonal blocks of $\Omega_A$
in which the number of particles in the subsystem A is restricted to $N_A$.
Correspondingly,
the eigenvalues $\lambda_\alpha$
of the full reduced density matrix $\Omega_A$
are grouped into those of $\Omega_{N_A}$'s.
Let us denote an eigenvalue of $\Omega_{N_A}$ as $\lambda^{(N_A)}_{\alpha}$.
Using
$\lambda^{(N_A)}_{\alpha}$,
we define $p_{N_A}$, then rescale the eigenvalues in this subspace as
\begin{eqnarray}
p_{N_A}&=&\sum_{\alpha} \lambda^{(N_A)}_{\alpha},
\label{pn}
\\
\tilde{\lambda}^{(N_A)}_{\alpha}&=&{1\over p_{N_A}} \lambda^{(N_A)}_{\alpha}.
\end{eqnarray}
Note that unlike in Eq. (\ref{S_A})
the summation over $\alpha$
in Eqs. (\ref{S_conf}) and (\ref{pn})
does not run over all the eigenstates of the full reduced density matrix $\Omega_A$
but only in the subspace with fixed $N_A$.

The number entropy $S_{\rm num}$ quantifies the fluctuation of
the number of particles in the subsystem A. 
Particle transport across the subsystems generates this type of entropy;
i.e., $S_{\rm num}$
is controlled by a {\it local} process.
Recent numerical studies have focused on whether or not 
$S_{\rm num}$ 
saturates in the MBL phase.\cite{Sirker,numberEELuitz,Zakrzewski}
If $S_{\rm num}$ 
is not saturated
in the putative MBL phase,
it implies that the particles are still transporting and in this sense
localization is not perfect.
%
The configuration entropy $S_{\rm conf}$
comes from {\it non-local} correlations in the many-body state,
while the distinction between
many-body and single-body localizations is also presumed to
be the presence or absence of non-local correlations.
Thus, the configuration entropy has the potentiality 
to probe 
such features that are specific to MBL, and
reveal 
its unique mechanism of localization;
\cite{DBHM,oritoconfig,QavityMBL} 
e.g., in Ref. \onlinecite{QavityMBL}, 
non-trivial entanglement growth due to cavity-mediated long-range interactions 
is evaluated via the use of 
configuration entropy.
%



\subsection{The entanglement entropy in the non-interacting limit:
wave packet spreading vs. entanglement entropy
}

Fig. \ref{fig_EE_V0}
shows time evolution of the entanglement entropy 
for the initial state (\ref{d_wave})
in non-interacting systems
with a variable strength of non-Hermiticity $g$.
Panels (i-a), (ii-a), (iii-a)
represent
the Hermitian case $g=0$;
the three panels represent, respectively,
(i-a) $S_{\rm tot}(t)$,
(ii-a) $S_{\rm num}(t)$,
(iii-a) $S_{\rm conf}(t)$.
The peculiarity of the Hermitian limit is
that having symmetric amplitudes in the two hopping directions,
the system is in a subtle equilibrium 
in which 
the interference of many plane-wave eigenstates
[cf. Eq. (\ref{psi_sup})]
leads to particle spreading in the delocalized regime.
%
Here,
in terms of the entanglement entropies,
they
show a rapid initial growth and a high saturated value
in such weak $W$ regime, while as 
$W$ is further increased to the localized regime:
the entropies tend to be strongly suppressed.
In panel (ii-a)
we focus on the behavior of the number entropy $S_{\rm num}(t)$.
In the regime of weak disorder: $W=0.4,1.6$, 
$S_{\rm num}(t)$ 
first rapidly grows,
and then saturates practically to the same value.
To be precise, as $W$ increases, 
the quick initial growth of $S_{\rm num}(t)$
tends to slow down toward the saturation
at the end of the initial growth.
In any case,
in the delocalized regime:
particles spread and 
after a sufficiently long time (although this time lengthens as $W$ increases)
the system reaches the equilibrium.
%
%
$S_{\rm conf}(t)$ also grows 
in the delocalized regime,
but 
the growth occurs with a slight delay 
with respect to that of
$S_{\rm num}(t)$ 
[see Fig. \ref{fig_EE_V0} (iii-a)]. 
As $W$ exceeds the critical value $W_c=2$,
both
the number and configuration entropies
decrease, but remain finite.
This is because
even in the localized phase, 
there is a finite localization length $\xi$,\cite{Sirker}
and especially
if $W$ is not too large and the system is not far from the transition,
this effect is non-negligible (see Appendix \ref{app_xi}).

We have seen in Sec. II-A
that
addition of non-Hermiticity $g\neq 0$ has a more drastic impact on 
the wave packet dynamics,
breaking the subtle equilibrium on the direction of particle motion,
and consequently,
the reality of the eigenvalues.
In the presence of $g\neq 0$
the wave packet ceases to spread, but starts to {\it slide}.
We have also seen in the previous section that
as $W$ increases,
this sliding velocity often increases, 
and when $W$ is further increased and approaches the critical value $W_c$,
the wave packet also starts to ``spread'', 
showing a {\it cascade-like} feature 
as in the hermitian case at weak $W$.
Under this circumstance
close to the localization transition
a particle 
may be rather undecided about which way to go 
in spite of 
the finite non-reciprocal hopping $g\neq 0$,
and
a situation somewhat similar to the Hermitian case
seems to be realized.
In the Hermitian case, we have seen 
that the entanglement entropies diminish as $W$ increases,
while in panels (i-b)-(i-d) of Fig. \ref{fig_EE_V0}, i.e., 
as $g$ is increased, this tendency starts to be uncertain and 
finally the tendency is just reversed. This result indicates that disorder generates entanglement entropy and showing a maximum at some finite $W=\tilde{W}_c$, then they turn to decrease.
\footnote{Number and configuration entropies also show similar behavior
[see e.g., panels (ii-b)-(ii-d) and (iii-b)-(iii-d) of Fig. \ref{fig_EE_V0}]}
We argue that
this maximum of the entanglement entropies
is related to the
delocalization-localization transition.
%
In Fig. \ref{fig_Smax} (a), 
the maximal value Max $S_{\rm tot}(t)$
\footnote{
Here, the maximal refers to
maximal in the time evolution 
for a given $g$ and $W$, unlike what the same word signifies in
a few sentences before.
}
of the entanglement entropy 
is plotted
in the ($W,g$)-plane.
The distribution of Max $S_{\rm tot}(t)$ resembles very much
that of $\sigma (t)=\Delta x (t)-\Delta x (0)$,  
in Fig. \ref{fig_1pv} (b),
showing a peak
close to the boarder Eq. (\ref{wc_NH2})
to the localized phase (if seen from the weak $W$ (delocalized) side);
i.e.,
``the maximum of Max $S_{\rm tot}(t)$'' coincides 
with the regime
where
a cascade-like expansion of wave packet occurs
in the density dynamics.
In other words,
the maximum of Max $S_{\rm tot}(t)$ occurs 
close to the delocalization-localization transition:
$\tilde{W}_c=W_c$.
The unusual 
behavior of entanglement entropies 
in the non-Hermitian non-reciprocal system
is thus revealed
to be directly related to
a very specific way how wave packet spreads (or rather, does {\it not} occur)
in this system.

Finally, let us comment on the number and configuration entropies. 
Figs. \ref{fig_EE_V0}(ii-b)-(ii-d) show the time evolution of number entropy,
in the delocalized phase, 
the number entropy first show a rapid initial growth with a speed almost independent of $W$,
then they show a second growth which looks more like a damped oscillation.
The mechanism of damped oscillation come from
the convergence to a single eigenstate $|\mu_0\rangle$ with the maximal imaginary part of eigenenergy $E_{\mu_0}$,
this also reflects the time evolution of density profile as shown in Fig. \ref{fig_damp}.
Convergence to a single eigenstate with the maximal imaginary part of 
eigenenergy also leads to suppression of the configuration entropy 
as shown in Figs. \ref{fig_EE_V0}(iii-b)-(iii-d). 
The convergence to a single {\it destined} eigenstate $|\mu_0\rangle$
implies 
the loss of superposition;
the configuration entropy is 
significantly reduced in this regime of $W\simeq 0$.
The destined
eigenstate is delocalized so that after certain time 
the initial density wave pattern Eq. (\ref{d_wall}) is almost washed out; 
see Fig. \ref{fig_damp},
while
the number and configuration entropies
show a damped oscillation
before they converge to a finial value.
Beyond a critical disorder strength $W_c$, the number, and configuration entropy 
look rather fluctuating (fast and randomly) as in the Hermitian case. 


\subsection{Case of $V\neq 0$ and $g\neq 0$ as well: the non-monotonic behavior}

We switch on the interaction: $V\neq 0$.
Similarly to the non-interacting case,
panels (i-a), (ii-a), (iii-a) of Fig. \ref{fig_EE_V}
show the evolution of the entanglement entropy
in the Hermitian limit: $g=0$;
the three panels represent, respectively, the cases of
(i-a) $S_{\rm tot}(t)$,
(ii-a) $S_{\rm num}(t)$,
(iii-a) $S_{\rm conf}(t)$.
Comparing these panels
with the corresponding panels in the non-interacting case [Fig. \ref{fig_EE_V0} (i-a), (ii-a), (iii-a)],
one can observe that
the total entanglement entropy $S_{\rm tot}(t)$ [panel (i-a)]
shows a quick initial growth until around 2-5$\times 10^1$, then
the growth slows down and smoothly
crosses over to a linear regime (in the logarithmic time scale).
Comparing these behaviors in panel (i-a) with the ones in panel (ii-a)
one notices that the quick initial growth comes from
the number entropy $S_{\rm num}(t)$,
while
the second slow growth
forming a linear regime 
stems from
the configuration entropy $S_{\rm conf}(t)$ in panel (iii-a).

As the non-Hermiticity $g$ is also switched on; see remaining panels of Fig. \ref{fig_EE_V},
the entanglement dynamics changes its face as in the non-interacting case;
cf. corresponding panels in Fig. \ref{fig_EE_V0},
but here, in the interacting case
a feature not existing in the non-interacting case
also come into play;
after the second slow growth 
the entanglement entropies 
start to decay,
exhibiting a characteristic
{\it non-monotonic} behavior. 
The non-monotonic behavior appears in the intermediate regime of $W$, but it is unclear whether it stems from the delocalized phase or localized phase. We employ the saturated value of the entropies to which are expected to be maximal at
\footnote{
especially, the magnitude of configuration entropy $S_{\rm conf}(t)$ [panel (iii-d)] 
is sensitive to the change of $W$.
}
$W_c$ close to the delocalization-localization transition as in the non-interacting case.
In Fig. \ref{fig_Smax}, panel (b),
the maximal value 
Max $S_{\rm tot}(t)$
in the time evolution 
is plotted in the parameter space $(W,g)$,
showing a broader maximum 
compared with the non-interacting case [panel (a)],
while the location of the peak
is slightly shifted to the side of larger $W$
from the non-interacting value (indicated by a white broken curve).
The location of the phase boundary 
estimated from the peak of Max $S_{\rm tot}(t)$ in Fig. \ref{fig_Smax} (b)
is consistent with an earlier numerical result:
$W_c\simeq 6-7$ at $g=0.5$ in Ref. \onlinecite{china}.
Comparing Fig. \ref{fig_Smax} with Fig. \ref{fig_EE_V},
one can observe that the non-monotonic behavior 
appears in the regime of $W$ corresponding to 
the critical-localized regime. 

The mechanism of 
the non-monotonic evolution,
which is a consequence of the competition between
(i) dephasing
and
(ii) convergence (or a gradual ``collapse'')
of the superposition [Eq. (\ref{Psi_sup})]
into a single (non-equilibrium steady) state \cite{panda}
$|\mu_0\rangle$ with a maximal imaginary part of the eigenenergy $E_{\mu_0}$.
In the absence of (ii),
(i) leads to
a slow but unbounded
(in the MBL phase, typically logarithmic)\cite{Abanin_log}
growth of the configuration entropy,
while
for (ii) to be operational 
the eigenenergy must have a finite imaginary part Im $E_\mu$.
Panels (iii-b)-(iii-d) of Fig. \ref{fig_EE_V} show 
the evolution of configuration entropy 
at different set of parameters.
They show that
in the localized regime
the non-monotonic behavior of the entanglement entropy 
(panels (i-b)-(i-d) of Fig. \ref{fig_EE_V})
stems mainly from the contribution of configuration entropy.
Note that
the dephasing enhanaces, while the convergence suppresses the configuration entropy.
In the localized regime these two effects compete, leading to a
non-monotonic entanglement evolution,
while in the delocalized phase, 
convergence dominates dephasing; therefore, no non-monotonic behavior appears.
In the non-interacting case: $V=0$,
the localization transition is believed to coincide with
the real-complex transition,\cite{Hatano1,Hatano2,Hatano3}
while
in the interacting case: $V\neq 0$,
whether 
the superposition (\ref{Psi_sup}) converges (collapses) to a non-equilibrium steady state
on the MBL side
is a more subtle issue.

Finally, we discuss the behavior of number entropy.
In the delocalized phase, 
the damped oscillatory behavior mentioned earlier is more conspicuous.
As the $W$ is increased, the damped oscillation disappears and 
replaced with a non-monotonic behavior with a broader maximum.
On top of the
non-monotonic behavior
one can also recognize a fast oscillatory component.
Such an oscillatory component
is also conspicuous
in the non-interacting case,
especially in the localized regime.
In the non-interacting limit
the disappearance of damped oscillation 
coincides with
the localization transition.
Here, we have shown that 
this is also the case
in an interacting system.

\begin{figure*}
\includegraphics[width=170mm]{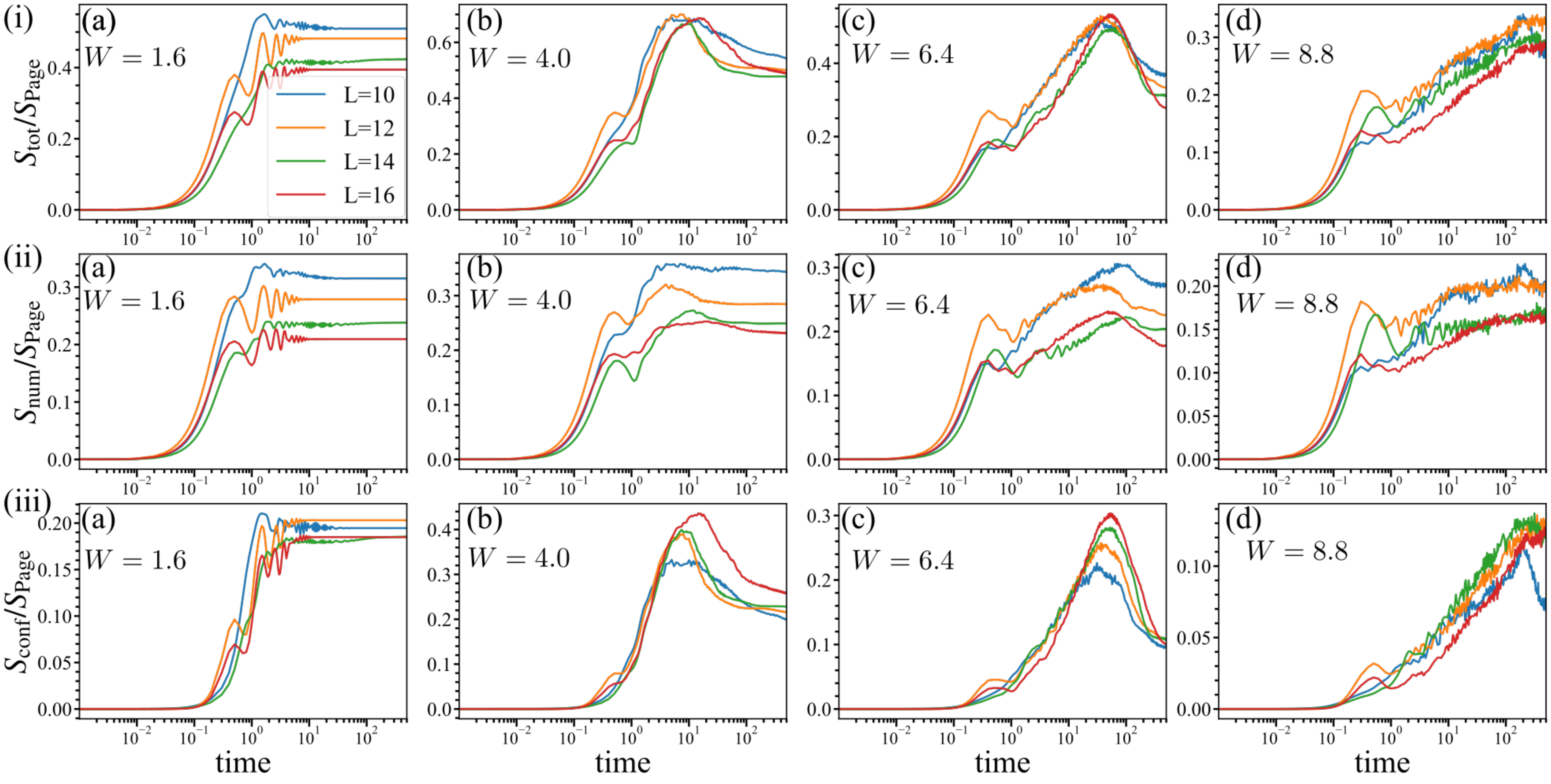}
\caption{
Time evolution of the entanglement entropies;
the first raw [panels (i-a)-(i-d)]: $S_{\rm tot} (t)$, 
the second raw [panels(ii-a)-(ii-d)]: $S_{\rm num} (t)$, and
the third raw [panels (iii-a)-(iii-d)]: $S_{\rm conf} (t)$.
The system is both
non-Hermitian: $g=0.6\neq 0$,
and interacting: $V=2$. 
Each column represents data at the same value of $W$, which evolves
from left to right as $W=1.6, 4.0, 6.4, 8.8$.
Curves plotted in different colors correspond to different system size,
as indicated in the inset of panel (i-a).
Density-wave type initial state (\ref{d_wave}); average over $50$ samples
with different $\theta_0$.
}
\label{fig_L1}
\end{figure*}

\begin{figure}
\includegraphics[width=80mm]{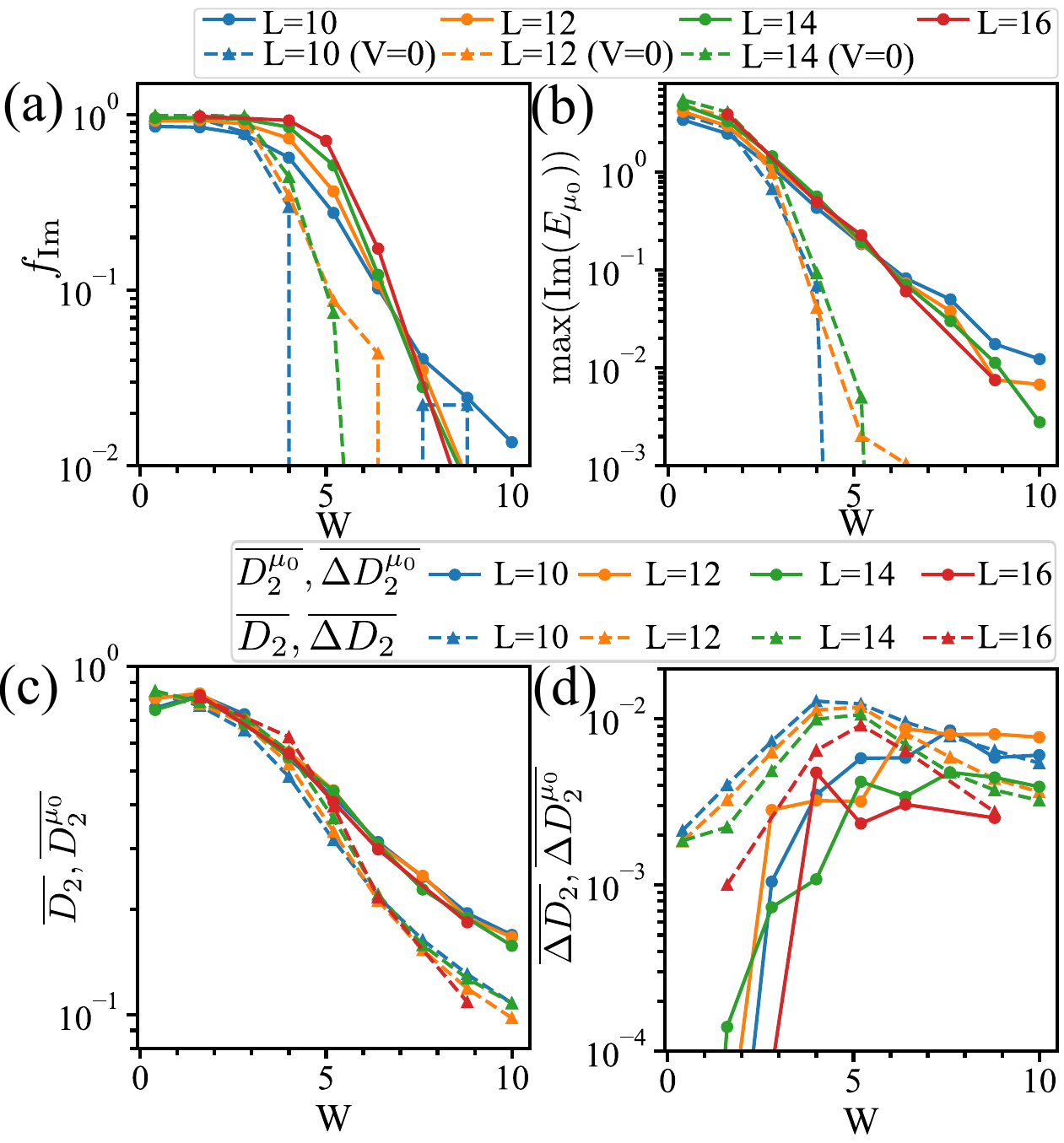}
\caption{Size dependence of
(a) the fraction of imaginary eigenvalues: $f_{\rm Im}$ [cf. Eq. (\ref{frac})],
(b) Max (Im $E_{\mu_0}$),
(c) the multifractal dimension $D_2$, and
(d) its fluctuation: $\Delta D_2$.
Data taken for an interacting case: $V=2$ 
(connected by {\it solid} lines)
is compared with the non-interacting case: $V=0$
(connected by {\it broken} lines).
Different colors correspond to different system size: $L=10, 12, 14, 16$.
$\theta_0$-averaged (50 samples).
}
\label{fig12}
\end{figure}

\section{Effects of finite size and non-equilibrium steady state}
\label{sectionIV}

Let us 
comment on the effects of system size.
In the three raws of 
panels in Fig. \ref{fig_L1}
we compare the evolution of the entanglement entropies:
(i-a)-(i-d) $S_{\rm tot} (t)$, 
(ii-a)-(ii-d) $S_{\rm num} (t)$, and
(iii-a)-(iii-d) $S_{\rm conf} (t)$ 
for systems of different system size $L$.
Note that the entropies are normalized by the ``Page value'',\cite{Pagesan}
given in the case of $N_A=N_B=L/2$
as 
\begin{equation}
S_{\rm Page}={L\over 2} \log 2-{1\over 2}.
\label{page}
\end{equation}
Note that $S_{\rm Page}$ is proportional to the size $L$ of the system;
if plotted curves; i.e.
the entanglement entropies calculated at different system size $L$
are size independent, then it means that
that entropy obeys the volume law
in the range of parameters in question.
In the Hermitian case
the magnitude of $S_{\rm tot} (t)$ is size independent\cite{Trans3,EEsys} 
at the stage of initial quick growth (area law),
while its saturated value after duration of a long enough time
is size-dependent ($\propto L$), obeying a volume law\cite{Abanin_log,BPM}.
In the first raw of Fig. \ref{fig_L1} 
[panels (i-a)-(i-d); different panels correspond to different strength of disorder $W$. $g$ is fixed at $g=0.6$.]
the total entanglement entropy $S_{\rm num} (t)/S_{\rm Page}$
is plotted at different system size $L$.
The magnitude of
$S_{\rm tot} (t)/S_{\rm Page}$ is clearly decreasing 
with the increase of $L$ in panel (i-a), implying an area-law behavior,
while 
in panel (i-b)-(i-c) 
[$W=4.0$ in panel (i-b), $W=6.4$ in panel (i-c); 
both values of $W$ are assumed to be not far from $W_c^{(V=2)}$]
it becomes size dependent at its maximum ($t\simeq 3-7\times 10^1$),
suggesting a volume-law behavior.
As $W$ is further increased [panel (i-d): $W=8.8$], 
the second growth of $S_{\rm num} (t)$ (a linear growth region in the logarithmic time scale)
lasts longer, and
the maximum of $S_{\rm num} (t)$ is not really achieved in the time scale
shown in the panel.

Comparing panels (i-a), (ii-a), (iii-a),
one can see that
the area-law size dependence of $S_{\rm tot}$ stems from
that of $S_{\rm num}$,
while
comparing panels (i-c), (ii-c), (iii-c),
one can see that
the volume-law behavior of Max $S_{\rm tot}$
results from an interplay of
the number and configuration entropies.
The insensitivity of 
Max $S_{\rm tot}$ to the system size $L$
suggests that
the non-monotonic evolution of the entanglement entropy 
specific to the non-Hermitian many-body system 
will also occur in cases of larger system size $L$.
%

For further clarifying this point 
we have plotted, in panel (a) of Fig. (\ref{fig12}), 
the ratio
\begin{equation}
f_{\rm Im}=D_{\rm Im}/D, 
\label{frac}
\end{equation}
where $D_{\rm Im}$
is the number of eigenenergies with a nonzero ($|{\rm Im} E_\mu|>10^{-13}$)
imaginary part Im $E_\mu$,
while
$D$ is the dimension of the Hilbert space;
i.e., the total number of eigenenergies. 
The existence of
nonzero Im $E_\mu$
leads to suppression of 
the superposition (\ref{Psi_sup}),
leading also to the suppression of
the entanglement entropies. 
The
delocalized eigenstates are susceptible of 
the non-reciprocity $g$ of hopping, 
generating a finite imaginary part in the eigenenergy.
Thus, the 
quantity $f_{\rm Im}$
measures the fraction 
of delocalized eigenstates in the total ensemble of eigenstates.
Panel (a) of Fig. \ref{fig12}
shows the evolution of the fraction
$f_{\rm Im}$ with the increase of $W$
for systems of different size $L$.
As $W$ is increased, $f_{\rm IM}$ first very gradually, then
quickly decreases;
this tendency 
is more accentuated
in systems of larger size $L$. 
Also,
the curves corresponding to different system size $L$
cross practically at the same point 
[at least in the interacting case: 
$V=2$, cases of solid curves in panel (a)]
at a value of $W=W_c\simeq 7$.
\footnote{In the non-interacting case $V=0$ [cases of dashed curves in Fig. \ref{fig12}, panel (a)]
the change of $f_{\rm Im} (W)$ becomes too drastic so that
one cannot really see the crossing itself,
but the overall behavior of $f_{\rm Im} (W)$ is similar to this interacting case.
}
In a system of size $L$ sufficiently large 
the curve $f_{\rm Im} (W)$ tends to become a sharply edged function:
$f_{\rm Im} (W)\simeq 1$ for $W<W_c$, while 
$f_{\rm Im} (W)\rightarrow 0$ for $W>W_c$;
i.e., in this case
practically all
the eigenstate, including highly excited states,
they altogether experience
a complex-to-real transition of eigenvalues;
they all, at least, most of them turn real at $W=W_c$.

In panel (b) of Fig. \ref{fig12}
we compare the (ensemble-averaged) 
maximal value of Im $E_\mu$
in the cases of $V=0$ and $V\neq 0$.
As expected, in the non-interacting case: $V=0$,
the magnitude of Max(Im $E_\mu$) experiences an abrupt fall
at  the crossing (size-independent) point:
$W=W_c\simeq 3-4$;
on the two sides of this transition point
the behavior of Max(Im $E_\mu$) shows a clear change 
that tends to magnify as the system size $L$ increases.
In particular, in the localized phase ($W>W_c$)
the magnitude of Im $E_\mu$ tends to vanish
as $L$ increases.
In the interacting case ($V\neq 0$) [Fig. \ref{fig12} (b) solid curves], 
on the other hand,
the ensemble-average of Max(Im $E_\mu$) 
does not show an abrupt fall;
instead, it gradually decays
over a broad range of $W$.
It shows no clear signature, 
in the ETH-MBL crossover/transition regime: $W\simeq W_c$.
Besides,
it shows no dependence on the system size $L$, either.
Based on these observations, we presume that
generically,
Im $E_\mu$ remains finite
even on the MBL side: $W> W_c$;
i.e.,
the superposition (\ref{Psi_sup}) converges to a non-equilibrium steady state
$|\mu\rangle=|\mu_0\rangle$ 
with a maximal Im $E_\mu$
even though it may take a longer time until the collapse
than in the ETH phase.
One may ask
what the nature of this steady state
$|\mu_0\rangle$ is.
Is it really a localized state?
This question arises, of course, since one usually associates 
the imaginary part of $E_\mu$
with an extended state.
%
To clarify this point
we have estimated the multi-fractal dimension 
\begin{equation}
D_2^\mu = -{\log(\sum_{\nu=1}^D |c_{\nu}^\mu|^4)\over \log(D)},
\label{D2}
\end{equation}
encoding the (de)localized nature of the eigenstate $|\mu\rangle$;
$D_2^\mu=1$ ($=0$) corresponds to a fully delocalized (localized) eigenstate.
$c_{\nu}^\mu$'s are the amplitudes of computational basis $|\nu\rangle=|n_1n_2\cdots n_L\rangle$ of $\mu$th eigenstate
and
$D$ is the dimension of the many-body Hilbert space.
 In panel (c) of Fig. \ref{fig12}, disorder averaged $\overline{D_2^{\mu_0}}$ (represented by solid lines) compared with disorder and all eigenstates averaged $\overline{D_2}$ (represented by broken lines).

The plots show that 
in the deep MBL regime, 
the eigenstates $|\mu_0\rangle$ show a localized tendency, 
even though its eigenenergy E has a finite imaginary part;
to be precise, the state $|\mu_0\rangle$ 
is less
localized than other eigenstates.
Panel (d) of Fig. \ref{fig12} shows
the variance $\overline{\Delta D_2}$ of the multi-fractal dimension 
as a function of $W$,
which suggests a qualitatively different behavior
in the ETH-MBL crossover regime
for
$\overline{\Delta D_2}$ and $\overline{\Delta D_2^{\mu_0}}$.
To summarize,
the superposition (\ref{Psi_sup}) 
indeed converges (collapses) to a non-equilibrium steady state
$|\mu_0\rangle$ 
even in the deep MBL phase,
while
the {\it destined} state $|\mu_0\rangle$ shows spatially a localized signature.

\begin{figure*}
\includegraphics[width=170mm]{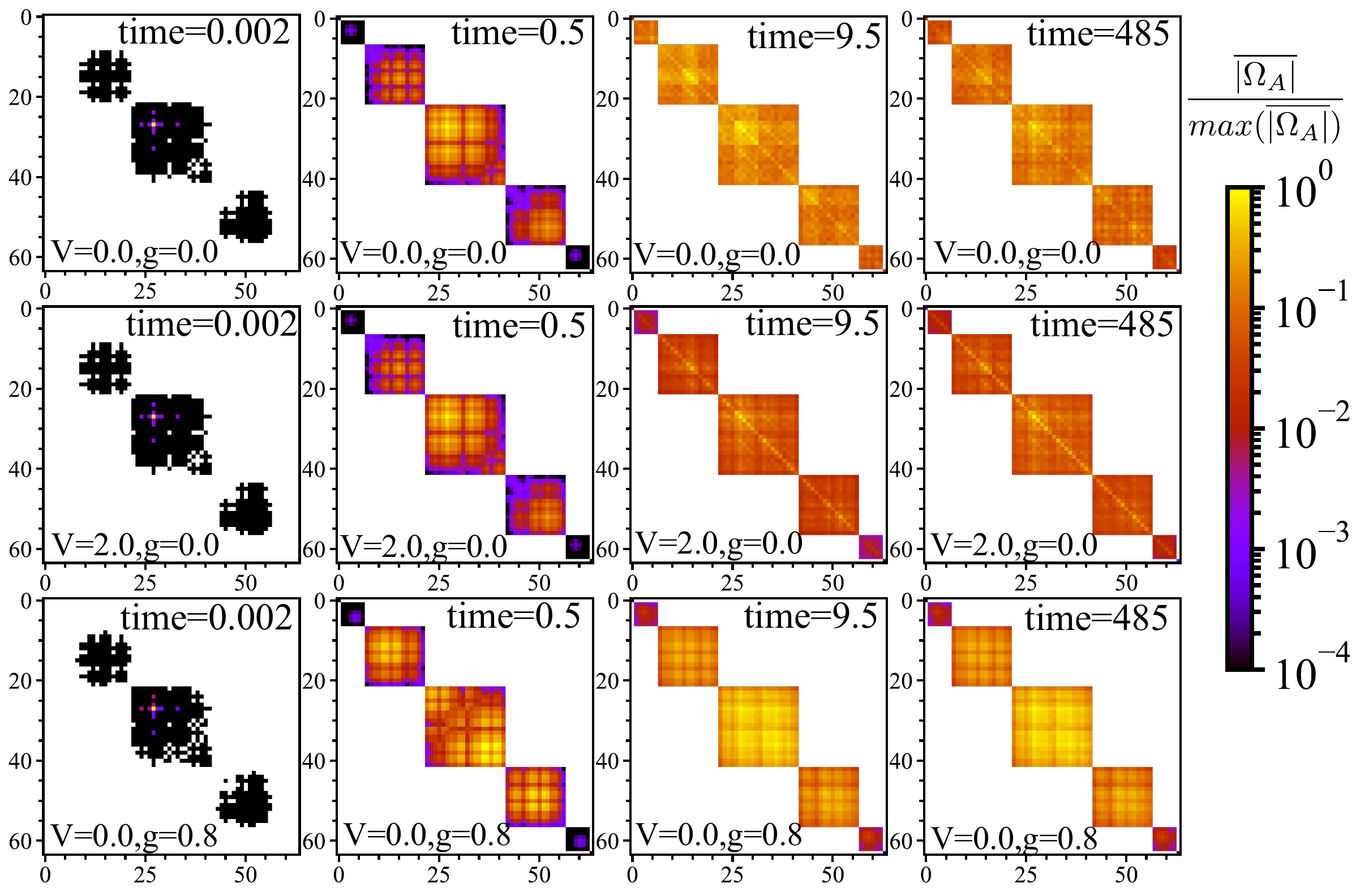}
\caption{
Evolution of the reduced density matrix in the wave-packet dynamics.
(i) the first raw: Hermitian limit ($g=0$), non-interacting case ($V=0$);
result of simple wave packet spreading.
(ii) the second raw: Hermitian limit ($g=0$), interacting case ($V\neq 0$);
both wave packet spreading and dephasing occur in
the dynamics.
(iii) the third raw: Non-Hermitian case ($g\neq 0$), non-interacting ($V=0$).
The same initial state as in Fig. \ref{fig_EE_V0};
$\theta_0$-averaged (50 samples).
}
\label{fig_RDM}
\end{figure*}

\section{Mapping the reduced density matrix $\Omega_A(t)$}
\label{sectionV}

Let us finally 
visualize
how different stages in the time evolution of the entanglement entropy can be understood
from the viewpoint of
the behavior of the reduced density matrix $\Omega_A(t)$.
In the twelve panels Fig. \ref{fig_RDM} 
we have plotted the elements of 
$\Omega_A(t)$
at different times
in order to visualize its time evolution.
%
We will see that
there are 4-5 characteristic stages in
the evolution of $\Omega_A(t)$. 

In the first raw of Fig. \ref{fig_RDM}, 
we start with the Hermitian case ($g=0$)
with no interaction ($V=0$).
At $t=0$
the initial state
$|\Psi (t)\rangle$ is chosen to be
a simple product state
so that
the reduced density matrix $\Omega (t=0)$
has a single finite element ($=1$)
in one of the diagonals.
As time passes by,
the region of a finite matrix element (represented by bright spots in the panels) 
spreads,
but for a while; e.g., case of the first panel from the left ($t=0.002$), 
most of the amplitudes
stay in the same diagonal block
in some $\Omega_{N_A}$ ($N_A=N_1,N_2,\cdots$) [cf. Eq. (\ref{RDM_bd})]
\footnote{
Recall that
the reduced density matrix $\Omega_A$
is block diagonalized as
in Eq. (\ref{RDM_bd}).
$\Omega_{N_A}$ represents a block of the reduced density matrix $\Omega_A$
such that
the number of particles in the subsystem A
is restricted to $N_A$.
}
From the start of the quench dynamics,
the system is susceptible to the process of wave packet spreading,
but
as far as $N_A$ is kept unchanged,
such a dynamics
does not lead to an immediate increase of the number entropy.
After some time; e.g., case of the second panel ($t=0.5$),
bright spots start to appear in the neighboring blocks,
and
finally; e.g., cases of the third and fourth panels ($t=9.5, 485$),
they become distributed almost equally in all the blocks.
It is expected that
the number entropy experiences
a rapid growth in this period and becomes saturated afterwards.

Let us switch on the interaction ($V\neq 0$)); 
Fig. \ref{fig_RDM}, the second raw.
In the first two panels,
the evolution of 
$\Omega_A (t)$
looks similar to the non-interacting case;
i.e., as time passes by,
bright spots tend to spread over all different blocks,
meanwhile
the number entropy tends to be saturated.
In the third and fourth panels,
the bright spots look
distributed equally in all the blocks
as in the non-interacting case,
but here
inside each block $\Omega_{N_A}$
they converge on the diagonals.
\footnote{
In the non-interacting case (first raw), on contrary,
in each submatrix $\Omega_{N_A}$
the bright spots look spread over all parts of the matrix; 
i.e., both in the diagonals and in the off-diagonals, almost randomly.
}
This is due to dephasing.\cite{Abanin_log}
Due to the interaction $V\neq 0$,
off-diagonal matrix elements
tend to acquire a random phase, which on average 
tends to vanish after some time.
Dephasing leads to increase of the configuration entropy 
$S_{\rm conf} (t)$;
thus,
in the present case
after the number entropy is saturated, 
the total entanglement entropy $S_{\rm tot} (t)$
continues to increase
with no systematic bound 
excepting the one due to the size of the system.

In the non-Hermitian case: $g\neq 0$ and also at $V=0$;
\footnote{
Here, we consider only the non-interacting case: $V=0$.}
Fig. \ref{fig_RDM}, the third raw,
the overall behavior of
the evolution of the reduced density matrix $\Omega_A (t)$
resembles 
the Hermitian case in the first raw
except that
here
inside each block $\Omega_{N_A}$
the pattern of the matrix elements
looks very regular,
while the patten looks random in the Hermitian case ($g=0$).
The reason will be the following:
here, in the regime of weak $W$,
the eigenenergy $E$ is typically complex,
so that
in the time evolution (\ref{Psi_sup}) of a (many-body) wave packet $|\Psi (t)\rangle$,
a single eigenstate $|\mu\rangle=|\mu_0\rangle$ with a maximal imaginary part ${\rm Im}\ E_\mu$
tends to predominate in the superposition (\ref{Psi_sup}).
We have already seen this tendency
in the evolution of the entanglement entropy $S_{\rm tot} (t)$ (cf. Fig. \ref{fig_EE_V0}); 
in the regime of weak $W$,
$S_{\rm tot} (t)$ quickly converges 
to a fixed value
and shows practically no fluctuation.

\section{Concluding remarks}
\label{sectionVI}

In this paper, we have elucidated the physics of non-Hermitian ETH-MBL transition 
from the viewpoint of
unusual (or rather, disappearance of)
wave packet spreading
and
through analyses of the time evolution of entanglement entropies.
In the first half of the manuscript,
we have highlighted
the nature of the unusual wave packet spreading
in a non-Hermitian system
with non-reciprocal hopping (Hatano-Nelson model) [Eqs. (\ref{ham_sp}), (\ref{ham_mp})].
Disorder has been modeled by a quasi-periodic potential
(Aubry-Andre model) [Eq. (\ref{W_qp})]. 
In a Hermitian system
wave packet spreading is
gradually suppressed by disorder, while
here in a non-Hermitian system 
it simply does not occur; 
even in the clean limit,
at least in the Hermitian way.
The wave packet rather {\it slides} and not spread.
Weak disorder exerts little effect on this behavior, while 
as it becomes strong enough 
to practically disable the non-reciprocity,
the characteristic sliding behavior tends to be replaced with
a cascade-like wave packet spreading analogous to the Hermitian case.
We have clearly demonstrated the mechanism [see Eq. (\ref{psi_gauss}), and related arguments] 
why the wave packet slides and not spread 
in our system,
and how such a behavior tends to be destroyed by the
quasi-periodic potential (cf. Fig. \ref{fig_psik}).
In the non-Hermitian case
the fundamental principle that governs the wave packet spreading
is altered
from the Hermitian case.

In the second half of the paper,
we have seen
how such unusual wave packet spreading in the non-Hermitian model
leads to
anomalous behaviors in the entanglement behavior. 
\footnote{As for another anomalous feature in the entanglement dynamics in a non-Hermitian PT model, see Refs. \onlinecite{ssonic1,ssonic2}.}
%
First, the specific wave packet spreading mentioned
leads to strong suppression of the entanglement entropy,
especially, in the delocalized phase
(and as the non-Hermiticity $g$ increases).
Second,
as turning on the interaction,
a finite imaginary part of the eigenenergy combined with the logarithmic growth (effect of dephasing)
leads to a characteristic non-monotonic behavior of the entanglement entropy. 
We have observed that
the maximum value of entanglement
in the time evolution becomes maximal at (or at least near) the 
delocalization-localization transition,
thus
containing information on the location of the transition.
This, in turn, signifies that the non-reciprocity $g$ can be used
as a probe for determining the localization length $\xi$ in the Hermitian system,\cite{Gil}
since
$g$ is directly related to $\xi$; i.e., $\xi=g^{-1}$ [see Eq. (\ref{xi_g}), and related footnote on it].
Note that 
identifying
the critical localization length $\xi_c$ is
a key to reveal the nature of the ETH-MBL transition.\cite{rare_thermal}
%
The size dependence of the entanglement dynamics shows
that
the entanglement entropy $S_{\rm tot} (t)$
at its maximum in the time evolution, i.e., Max $S_{\rm tot} (t)$
obeys the volume law, while
under other circumstances (typically, in the delocalized phase)
$S_{\rm tot} (t)/S_{\rm Page}$ tends to decrease with the increase of $L$,
suggesting an area law.
Thus,
the entanglement entropy in a non-Hermitian system
exhibits
an unusual area-volume-area law type crossover
as a result
of the interplay between the
non-Hermiticity and the interaction.
%

As pointed out in Ref. \onlinecite{panda},
collapse of the superposition in the initial state
and convergence
to a single destined eigenstate $|\mu_0\rangle$ with maximal Im $E_{\mu_0}$
play a central role in the behavior of the entanglement entropy in this non-Hermitian system.
Here,
we have verified this point
on a more firm basis 
through analyses 
of the entanglement 
using different indices (e.g., number and configuration entropies)
and
in a wide range of parameter regimes.
On the localized side,
most of the eigenstates are localized, 
but 
a few continues to have
a small but finite imaginary part Im $E$ 
enough to compete with the logarithmic growth (dephasing) 
at least at a very long time scale,
leading to the characteristic non-monotonic behavior of $S_{\rm tot} (t)$.

Finally,
it will be challenging 
to extend the analyses done in this work in systems of larger size. 
In Hermitian systems, 
Krylov-based time evolution\cite{Trans3} and tensor network techniques\cite{tensornet1,tensornet2,tensornet3}
are known to be applicable to deal with systems of large size.
The extension of these techniques to the non-Hermitian case
will be a possible direction to proceed in a future work.

\acknowledgments
We are grateful to Naomichi Hatano, Tomi Ohtsuki, Ivan Khaymovich and Takeshi Okayasu
for useful comments and discussions.
We have employed QuSpin\cite{Quspin1,Quspin2} 
for generating the explicit matrix elements of
the Hamiltonian 
such as the ones given in Eqs. (\ref{ham_sp}), (\ref{ham_mp}).
KI has been supported by JSPS KAKENHI Grant Number
21H0100501, 20K037880 and 18H0368352.

\appendix

\section{Notes on the choice of the boundary condition}
\label{app_BC}

The non-Hermitian system with non-reciprocal hopping
is extremely sensitive to the choice of the boundary condition,
at least in its statics.
Under the open boundary condition (OBC)
the system exhibits a so-called skin effect;\cite{YW}
all the bulk wave functions
become actually {\it skin modes} in this case, while 
the complex spectrum is unique to the periodic boundary condition (PBC);
all the eigenvalues are {\it real} under the OBC.

The sensitivity of the system
to the boundary condition has already been recognized
in the original works of Hatano and Nelson, \cite{Hatano1,Hatano2}
but recent intensive discussions on non-Hermitian topological insulators
have revealed its even further consequences.
In topological insulators, 
the bulk topology under the PBC
is in one-to-one correspondence
with
the appearance/absence 
of edge states under the OBC 
(the bulk-edge correspondence),
while in the non-Hermitian system in question
the eigen wave functions tend to damp or promote spatially under the OBC;
they become skin modes.
Spatially, they behave like edge modes, but 
in a sense they should be rather regarded as bulk states
(a confusing situation),\cite{YM}
and what is worse,
they are incompatible with the PBC,\cite{imura1,imura2} 
which one usually applies
in the bulk geometry, etc.
In short, a strong dependence of the system's static properties
on the boundary condition is an obstacle
for applying the concept of Hermitian topological insulator
to this system.
As another remark,
the above skin effect is proposed to be in itself topological;
its occurrence (under the OBC) 
is protected by a specific winding property of
the (complex) spectrum in the complex energy plane
under the PBC.
\cite{okuma}

Under the PBC, on the other hand,
the eigenenergies of the system take complex values;
especially, the system exhibits a complex spectrum
in the clean limit.
The complex nature of the spectrum (Im $E\neq 0$) is closely related to the
plane wave nature of the eigen wave function;
under the OBC this is no longer true even in the clean limit.
In the presence of weak disorder
the eigen wave functions are extended as far as it is not too strong
and that is just enough for
keeping the spectrum complex.
Too strong disorder makes the eigen wave function localized,
pushing the spectrum down to the real axis (if one focuses on what happens in the upper-half complex plain).
Thus,
the delocalization-localization transition in this system
is simultaneously
a spectral transition from complex to real.
In an interacting system: $V\neq 0$,
one can expect that this still holds.
Indeed, 
based on the observation that
numerically estimated critical points of
the ETH-MBL and complex-real transitions are close (not incompatible),
also reinforced by some analytic arguments,
the authors of Ref. \onlinecite{Hamaz}
{\it conjecture}
that the two transitions actually coincide in the thermodynamic limit.
The authors of Ref. \onlinecite{china},
on the other hand, 
add to the (conjectured) {\it double transition} 
the third topological transition mentioned in the last paragraph;
i.e.,
at the double transition,
the winding property of the complex spectrum under the PBC
also changes,
which is in one-to-one correspondence with the disappearance of
the non-Hermitian skin effect under the OBC.

\begin{figure}
\includegraphics[width=50mm]{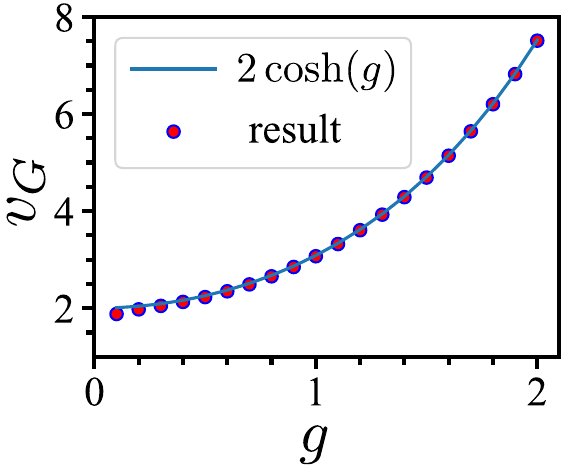}
\caption{$g$ dependence of the sliding velocity $v_G=[x_G(t)-x_G(0)]/t$
in free particle dynamics: $W=0$;
cf. Fig. \ref{fig_1pv} (a).
From the same data as employed for producing 
Fig. \ref{fig_1pv} (a)
data at $W=0$ is extracted and replotted as a function of $g$.
To be compared with the formula (\ref{vg}). 
No disorder averaging ($\theta_0=0$); $j_0=580$. $L=601$.}
\label{fig_append}
\end{figure}

\section{Free particle wave-packet dynamics 
in a non-Hermitian system with Im $\epsilon_k\neq 0$}
\label{app_gauss}



Let us consider a free particle motion
prescribed by the Hamiltonian (\ref{ham_sp}). 
For simplicity,
we switch off the quasi-periodic potential (\ref{W_qp}); $W=0$.
In this {\it disorder-free} case,
the eigenstates of the Hamiltonian (\ref{ham_sp}) under the PBC
take the form of a plane wave $e^{ik}$, 
while the corresponding eigenenergies $\epsilon_k$ become
{\it complex}; see Eqs. (\ref{spec_comp}), (\ref{spec_ellip}).
Taking the initial state $|\psi(t=0)\rangle$ as in the main text; i.e., as in Eq. (\ref{psi_0}), 
we consider its time evolution Eq. (\ref{psi_sup}).
Considering
the limit $L\rightarrow\infty$,
we replace the summation over $k$ in the last line of Eq. (\ref{psi_sup})
by an integral:
\begin{equation}
|\psi(t)\rangle=\sum_{j}|j\rangle\int^{2\pi}_{0} dk \frac{1}{\sqrt{2\pi}}e^{2i\cos(k-ig)t+ik(j_0-j)}.
\label{sup_k_int}
\end{equation}
At this point, let us recall that
the one-body spectrum $\epsilon=\epsilon_k$ is complex as in Eq. (\ref{spec_comp}), and
its imaginary part Im $\epsilon_k$ becomes maximal at $k=k_0=-\frac{\pi}{2}$.
Then,
in the superposition of contributions from different $k$-components
(the integral over $k$)
in Eq. (\ref{sup_k_int}), 
the dominant contributions in time evolution
stem from those around $k=k_0$,
and such contributions govern the long time dynamics;
\begin{eqnarray}
|\psi(t)\rangle
&\simeq&\sum_{j}|j\rangle\int^{-\frac{\pi}{2}+\delta k}_{-\frac{\pi}{2}-\delta k} dk \frac{1}{\sqrt{2\pi}}e^{2i\cosh(g)(k+\frac{\pi}{2})t}\nonumber\\
&&\times e^{2\sinh(g)(1-\frac{1}{2}(k+\frac{\pi}{2})^2)t+i(k+\frac{\pi}{2})(j_0-j)}\nonumber\\
&\simeq&\sum_{j}|j\rangle\int^{\infty}_{-\infty} dk \frac{1}{\sqrt{2\pi}}e^{2i\cosh(g)kt}\nonumber\\
&&\times e^{2\sinh(g)(1-\frac{k^2}{2})t+ik(j_0-j)}\nonumber\\
&=&\sum_{j}|j\rangle \exp(-\frac{((j_0-j)+2\cosh(g)t)^2}{4\sinh(g)t})\nonumber\\
&&\times e^{2\sinh(g)t}/\sqrt{4\sinh(g)t}.
\label{psi_gauss_app}
\end{eqnarray}
From the last expression, one can read (the $g$-dependence of) the group velocity 
of the wave packet as
\begin{equation}
v_G=2\cosh(g), 
\label{vg}
\end{equation}
which seems also consistent with our numerics
(Fig. \ref{fig_append}).
Note that $v_G$ increases with the increase of $g$.
The last expression
takes the form of a 
Gaussian wave packet
that {\it slides} in the direction imposed by $g$,
clearly demonstrating
why the packet does not immediately spread, but rather {\it slides}.
The expanse $\Delta x$ of the wave packet
gradually increases in real space, though,
in the course of time as
$\Delta x \simeq 2\sqrt{\sinh(g)t}$,
while
the corresponding width
$\Delta k\sim 1/\Delta x$
in the reciprocal space tends to diminish [see Fig. \ref{fig_psik} (a)].
The above features are also quite manifest 
in panels (a-b) of Fig. \ref{fig_1pd} 
albeit in the panels
a weak disorder is present.
These are consequences of
the fact that 
in the time evolution
the state $|\psi(t)\rangle$ as given in Eq. (\ref{sup_k_int}), 
tends to be governed by
the eigenstates with maximal Im $\epsilon_k$ [see Eq. (\ref{psi_gauss})],
while
the individual eigenstates 
take spatially a form of 
plane wave.

Note that
the form of the Gaussian wave packet (\ref{psi_gauss_app})
suggests that
it obeys a square-root scaling
typical to classical diffusion dynamics.
Of course,
we consider a coherent quantum dynamics of
the wave function $\psi(x,t)$, which is governed by the
Schr\"odinger equation.
Here, however,
in the present non-Hermitian setup, 
the asymmetric nature of the hopping strongly suppresses
the interference of the complex wave function $\psi(x,t)$,
characteristic to the Schr\"odinger quantum dynamics.
As a result,
the wave function $\psi(x,t)$ as given in Eq. (\ref{psi_gauss_app})
obeys effectively
the classical diffusion dynamics.
In the presence of random potential scatterers,
different scattering paths are expected to interfere in spite of the asymmetric hopping;
i.e., quantum interference may be partly recovered in this case.
As a result, the system may exhibit a behavior analogous to the Hermitian case.
Such an expectation is 
consistent with the appearance of the cascade-like enhancement of the 
wave-packet spreading in the regime of $W$ close to $W_c$
(see Sec. II).

\begin{figure}
\includegraphics[width=60mm]{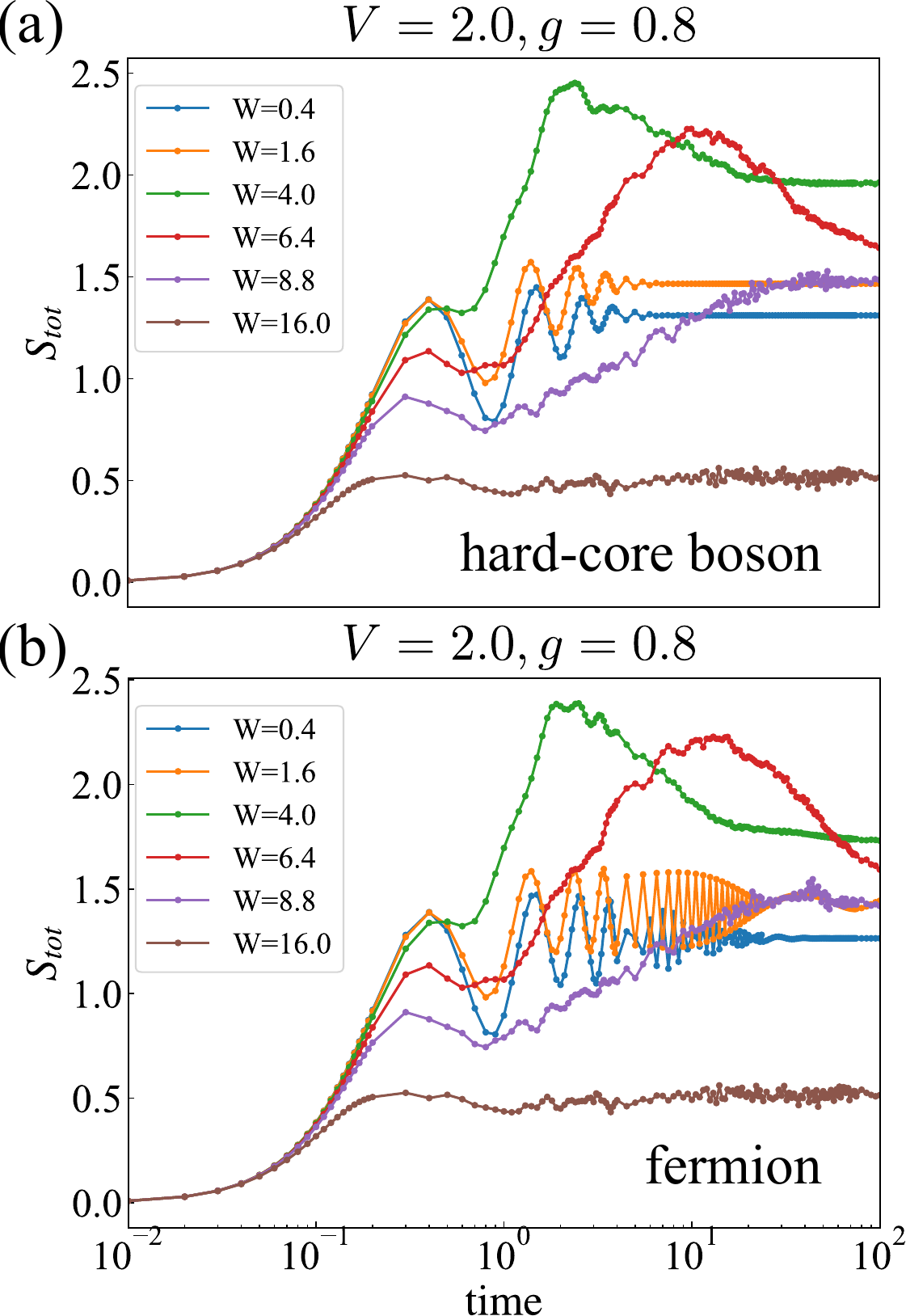}
\caption{Time evolution of the entanglement entropy
in the (a) hard-core boson and (b) fermion models.
We employ same numerical parameters as in Fig. \ref{fig_EE_V}.}
\label{fig_appendC}
\end{figure}
\section{Effects of anticommutation relation on the entanglement dynamics}
\label{app_fermi}

So far we have considered systems of hard-core bosons;
cf. Refs. \onlinecite{Hamaz}, \onlinecite{china}.
In a system of true fermions,
the anticommutation of $b_j$ and $b_j^\dagger$
gives rise to a subtle sign difference 
at the periodic boundary.
Here, we examine whether
this sign difference results in 
a significant consequence in the entanglement dynamics.
Fig. \ref{fig_appendC}(a) shows selected plots from Fig. \ref{fig_EE_V} (i-b), 
which demonstrates time evolution of the entanglement entropy 
in the case of $g=0.8$ in the hard-core boson model.
Fig. \ref{fig_appendC}(b) represents corresponding data 
in the fermion model; under the same condition. 
In the critical and localized regimes: $W>3-4$,
the wave functions are not extended,
and the corresponding electronic states
are expected to be immune to the sign difference.
The result of our simulation in the fermionic case
confirms that the entanglement entropy exhibits 
essentially the same behavior in the two models; 
i.e., the anticommutation relation plays no role in this regime.
In the delocalized regime ($W<3-4$) 
the wave functions are extended over the entire system so that,
in principle, the resulting electronic state 
may be influenced by the statistics.
In this regime the entanglement entropy shows
an oscillatory behavior in the intermediate time scale;
a simple damped oscillation in the case of
hard-core boson model. In the fermionic case one can
still recognize a similar oscillatory pattern,
but it does not look any longer a simple damped oscillation,
and also survives longer than in the hard-core boson case.
Thus, the anticommutation relation influences the behavior
of the entanglement entropy in the delocalized regime
at a quantitative level, but the qualitative statements
on its behavior in the hard-core boson model can still
be applicable to the fermionic case. The behavior of the
entanglement entropy is essentially unchanged in the critical and localized regimes.

\section{The entanglement entropy in the ETH and MBL phases (Hermitian case)}
\label{app_xi}

Let us comment on the relation
between the entanglement entropy and the localization length. 
The multifractal dimension $D_2$ given in Eq. (\ref{D2})
measures 
to what extent
the eigenstates are localized 
in the Hilbert space;
this is related to how much 
the wave functions are localized in real space.
Thus,
in the regime of strong disorder
(and in the Hermitian case: $g=0$),
$D_2$ behaves asymptotically as \cite{rare_thermal}
$D_2\sim\log(\xi)$,
where
$\xi$ represents the localization length in real space. 
This implies that
the true Fock space-localization $D_2=0$
is generally 
never achieved \cite{FSLaf,FSmil,FockLO},
since $\xi$ is generally 
finite in the localized phase,
except in the limit $W\rightarrow\infty$.

Fock-space localization 
effectively restricts the available Hilbert space, thus
affecting the maximal value of the entanglement entropy;
cf. in the free case
it is given by the Page value (\ref{page}).
In a more generic case with a finite $D_2$
it becomes,\cite{MFDmeetEE}
\begin{equation}
S_{\rm tot}\sim D_{\rm ent}{L\over 2}\log(2),
\label{D2EE}
\end{equation}
where
$D_{\rm ent}$ is a quantity related to $D_2$:
\begin{equation}
D_{\rm ent}=
\begin{cases}
    1, & D_2 \ge 1/2 \\
    2D_2,        & D_2 < 1/2.
\end{cases}
\label{Dent}
\end{equation}
i.e.,
on the ETH side: $D_2>1/2$
the reduction of the multifractal dimension $D_2$ due to
a finite $W$
does not lead to reduction of the entanglement entropy $S_{\rm tot}$,
while
in the MBL side: $D_2<1/2$,
$S_{\rm tot}$ decreases linearly with the decrease of $D_2$;
in particular,
$S_{\rm tot}$
does not show a finite jump
at the ETH-MBL transition: $D_2=1/2$,
and gradually crosses over from the ETH to the MBL side.
In the time evolution of $S_{\rm tot} (t)$
plotted in Fig. \ref{fig_EE_V0} (a),
its saturated value in the long time scale
remains the same in the delocalized phase: $W\le2$, while
in the localized phase: $W>2$,
it decreases continuously from this value
as $W$ is increased.

\bibliography{ref_MBL_NH7}

\end{document}